\documentclass[letterpaper,11pt]{article}
\usepackage[utf8]{inputenc}

\usepackage{jheppub}
\usepackage{graphicx,verbatim}
\usepackage{blindtext}
\usepackage[T1]{fontenc}
\usepackage{epsf}
\usepackage{lmodern}
\usepackage[mathscr]{euscript}

\usepackage{ytableau}
\usepackage{youngtab}

\usepackage{amsmath}
\usepackage{amsfonts}
\usepackage{amssymb}
\usepackage{mathrsfs}
\usepackage{slashed}
\usepackage{xcolor}
\usepackage{tikz, float}
\usetikzlibrary{patterns,shapes.misc}
\usepackage{circuitikz}

\input{preambles.sty}

\def\fD{\mathfrak{D}}

\def\BC{\mathbb{C}}

\def\BR{\mathbb{R}}

\def\BZ{\mathbb{Z}}

\def\CalN{\mathcal{N}}

\def\CalO{\mathcal{O}}

\def\CalQ{\mathcal{Q}}

\def\CalX{\mathcal{X}}
\def\Tr{{\rm Tr}}

\def\ve{{\varepsilon}}

\def\fg{\mathfrak{g}}

 \def\a{\alpha}
 \def\b{\beta}
 \def\g{\gamma}
 \def\d{\delta}

 \def\k{\kappa}
 
 \def\m{\mu}

 \def\s{\sigma}
 \def\t{\tau}

 \def\G{\Gamma}
 \def\D{\Delta}

 \def\S{\Sigma}
 \def\L{\Lambda}

\def\beq{\begin{equation}}
\def\eeq{\end{equation}}

\title{Dimers for Relativistic Toda Models with Reflective Boundaries}
\author[a]{Kimyeong Lee}  
\author[b]{and Norton Lee}
\affiliation[a]{Beijing Institute of Mathematical Sciences and Applications (BIMSA), Huaibei Town, Huairou
District, Beijing 101408, China}
\affiliation[b]{Center for Geometry and Physics, Institute for Basic Science (IBS), \\Pohang 37673, Korea}
\emailAdd{klee@bimsa.cn, norton.lee@ibs.re.kr}

\preprint{CGP25010}

\vspace{2cm}
\abstract{We construct dimer graphs for relativistic Toda chains associated with classical untwisted Lie algebras of A, B, C$_0$, C$_\pi$, D types and twisted A, D types. We show that the Seiberg-Witten curve of 5d $\CalN=1$ pure supersymmetric gauge theory of gauge group $G$ is a spectral curve of the relativistic Toda chain of the dual group $G^\vee$. {}  }

\begin{document}

\maketitle

\section{Introduction}

N. Seiberg and E. Witten proposed their ansatz \cite{Seiberg:1994aj} for the 4d $\CalN=2$ supersymmetric gauge theories. Since then there have been many attempts to realize the structure behind it, in order to have better understanding. 
One of the important structures of the Seiberg-Witten (SW) ansatz is integrability. Concretely \cite{Gorsky:1995zq,Martinec:1995by}, it has been shown that the dynamics of the pure supersymmetric gauge theory with the gauge group $G$ can be described in the framework of periodic Toda chain for the dual group $G^\vee$, whose affine Dynkin diagram is dual to that of $G$. 
The Seiberg-Witten curve of supersymmetric gauge theory coincides with the spectral curve of the Toda chain. 
Since then, there has been a lot of work exploring the correspondence between the SW curve and integrable systems \cite{Marshakov:1997cj,Gorsky:1996qp,Nekrasov:1996cz,Donagi:1995cf,Gorsky:1997mw,Marshakov:2012kv,Haouzi:2020yxy}. The list includes theories in 4d, 5d, and 6d with matter hypermultiplets in the adjoint or fundamental representations. 
The integrable systems in correspondence to the 5d $\CalN=1$ supersymmetric gauge theories are often relativistic. 

Relativistic Toda chains (RTCs for short) were originally introduced in \cite{Ruijsenaars_RToda1990}. For this Toda chain, which is naturally associated to the root system $A_{N-1}$,  was first solved with the help of $N \times N$ Lax matrix formalism, later by the $2 \times 2$ Lax operator formalism that obeys the Sklyanin's quadratic algebra. The two different Lax formalism are equivalent and can be converted to one another using the Baker-Akhiezer function. 

E. Sklyanin \cite{Sklyanin:1988yz} points out that the $2 \times 2$ Lax formalism is better suited for the Toda lattices defined based on other classical Lie algebra. Toda lattice defined on the root systems of type B, C$_0$, C$_\pi$, and D are described by change of boundary conditions for the ordinary (type A) Toda lattice in accordance to the structure of the Dynkin diagram of the Lie algebra $\fg$. 
The reflection matrix, representing the change of boundary conditions in a spin chain, is constructed in \cite{Sklyanin:1988yz,Gorsky:1999gx} for non-relativistic one, and relativistic chain in \cite{Kuznetsov:1994ur}. 

The RTCs arises naturally from the Lie group \cite{Fock:1997aia,Kruglinskaya:2014pza} and thus have a cluster structure. The cluster integrable systems are relativistic integrable systems \cite{Kenyon2011dimers,Marshakov:2012kv,Fock:2014ifa} with the log-canonical Poisson structure encoded in a quiver $\CalQ$, which is identical to the (point-like) BPS quiver of the 5d supersymmetric gauge theory compactified on a circle \cite{Closset:2019juk}. 

The dual graph of a quiver $\CalQ$ is a planar, periodic dimer graph $\G$ on $T^2$. The spectral curve of the cluster integrable are obtained from the \emph{Kasteleyn matrix} $\fD$, a weighted adjacency matrix of the dimer graph $\G$. The spectral curve of the cluster integrable system defined on a dimer graph $\G$ is 
\begin{align}
    \det \fD(x,y) = 0~.
\end{align}
Alternatively, the same spectral curve can be obtained from the $2 \times 2$ Lax operators, which comes from affine Lie group construction. This identifies the cluster variety with a Poisson submanifold in the co-extended affine group.

The dimer graph for RTC associated with the root system $A_{N-1}$ is well studied \cite{Hanany:2005ve,Franco:2005rj,Eager:2011dp,Lee:2023wbf}. The RTCs associated with the other classical Lie algebra are less understood. 
In \cite{Lee:2024bqg} the dimer graph for type D RTCs are constructed. 
We want to complete the story by constructing dimer graph for RTCs defined on all classical Lie algebra (except the exceptional Lie algebras) and their twisted variants.

\paragraph{Outline}

We start with a review on the integrability of relativistic Toda chains (RTC) in the Lax formalism in section~\ref{sec:RTC}. We will review on E. Sklyanin's description of RTC defined by Lie algebras of B, C, and D type in section~\ref{sec:RTC-bdy}. They are considered as type A RTC with reflective boundary condition. 
In section~\ref{sec:dimer} we review on how the Lax operators naturally arise from the Kasteleyn matrix of the cluster integrable system for type A RTC. 
Then we demonstrate how the reflection matrices representing the reflective boundary for type B, C, and D RTC also arise from specific modification of a dimer graph. 
In section~\ref{sec:example} we construct dimer graph for RTC defined on various untwist and twisted Lie algebra. 
Finally we point out our summary and potential furture direction in section \ref{sec:summary}.

\paragraph{Acknowledge}
The authors thank Mohammad Akhond, Saebyeok Jeong, Minsung Kim, Minsung Kho, Rak-Kyeong Seong for fruitful discussion and correspondence. The work of NL is supported by the IBS project IBS-R003-D1. The work of K.L. is supported in part by the Beijing Natural Science Foundation International Scientist Project (No.:IS25024) and BIMSA start-up fund.

\section{Relativistic Toda chains}\label{sec:RTC}

It is well known that the Lax operator of the Toda chain can be constructed from the Heisenberg XXX magnet \cite{Gorsky:1995zq,Gorsky:1997jq}. 
For the Heisenberg XXZ magnet, one can take the analogous contraction and construct the Lax matrix of the form
\begin{align}\label{def:Lax}
    L(x;q_n,p_n) = \begin{pmatrix}
        x^{\frac12}e^{-\frac{p_n}{2}} - x^{-\frac12}e^{\frac{p_n}{2}} & -\L e^{q_n} \\ \L e^{-q_n} & 0
    \end{pmatrix}
\end{align} 
where $p_n$ and $q_n$ are the conjugate momentum and position of the $n$-th particle. $\L$ is the coupling constants of the RTC. 

The Lax matrices $L(x;q_n,p_n)$ obey the Sklyanin's quadratic algebra, also known as the \emph{RLL relation}: 
\begin{align}\label{eq:Sklyanin-qua}
    \left\{ \overset{1}{L}(x),\overset{2}{L}(x') \right\} = \left[ r(x/x'), \overset{1}{L}(x) \overset{2}{L}(x') \right]
\end{align}
where $\overset{1}{L}(x) = L(x) \otimes I$ and $\overset{2}{L}(x) = I \otimes L(x)$. 
The classical trigonometric \emph{R-matrix}
is given by 
\begin{align}
    r(x) = \begin{pmatrix}
        \frac12 \frac{x^2+1}{x^2-1} & & & \\
        & -\frac12 & \frac{x}{x^2-1} & \\
        & \frac{x}{x^2-1} & \frac12 &   \\
        & & & \frac12 \frac{x^2+1}{x^2-1}
    \end{pmatrix}. 
\end{align}
The $2\times 2$ monodromy matrix $t_N(x)$ of $A_N^{(1)}$ Lie algebra RTC is given by the congregation of the Lax matrices 
\begin{align}
    t_N(x) = L(x;q_N,p_N) \cdots L(x;q_1,p_1). 
\end{align}
It is obvious that by definition the monodromy matrix $T(x)$ obeys the same Sklyanin's quadratic algebra \eqref{eq:Sklyanin-qua} as the Lax matrices. 
The spectral curve of $A_N$ RTC is 
\begin{align}\label{eq:spec-A}
    y + \frac{\det t_N(x)}{y} = \Tr~ t_N(x) = e^{-\frac{p_1+\cdots+p_N}{2}} \left( x^{\frac{N}{2}} + \sum_{n=1}^N (-1)^nH_n x^{\frac{N}{2}-n} \right). 
\end{align}
Here $H_n$, $n=1,\dots,N$, are the conserving Hamiltonians, obeying 
\begin{align}
    \{H_n,H_m\} = 0~,~ n,m=1,\dots,N. 
\end{align}
The Hamiltonian of type A RTC is
\begin{align}
    H_1 = \sum_{n=1}^N e^{p_n} + \sum_{n=1}^{N-1} \L^2 e^{q_{n+1}-q_n} e^{\frac{p_{n+1}+p_n}{2}} + \L^2 e^{q_1-q_N}e^{\frac{p_1+p_N}{2}}~.
\end{align}



\subsection{RTC with reflective boundaries}\label{sec:RTC-bdy}

Observed by Sklyanin, RTCs of type B,C, and D Lie algebras can be viewed as type A RTC with reflective boundary conditions.  
The monodromy matrix of RTC with reflective boundary conditions is given by
\begin{align}\label{eq:monodromy-reflection}
\begin{split}
    T(x) = & ~ K_+(x) t_N(x) K_-(x) t^-_N(x). 
\end{split}
\end{align}
where $t_N(x)$ is the monodromy matrix of $N$ particle type A RTC, and 
\begin{align}\label{def:Lax-2}
\begin{split}
    t_N^-(x) & = \det t_N(x) \times t_N(x^{-1})^{-1} = \s t^T_N(x^{-1}) \s^{-1}~,~ \s = \begin{pmatrix}
        0 & 1 \\ -1 & 0
    \end{pmatrix}~. \\
\end{split}
\end{align}
To keep the system integrable, the reflection matrices $K_\pm(x)$ must obey the \emph{reflection equation}: 
\begin{subequations}\label{eq:reflect-classical}
\begin{align}
    \left\{ \overset{1}{K_\pm}(x),\overset{2}{K_\pm}(x') \right\} = &\left[ r(x/x'), \overset{1}{K_\pm}(x) \overset{2}{K_\pm}(x) \right] \\
    & + \overset{1}{K_\pm}(x) r(xx') \overset{2}{K_\pm}(x') - \overset{2}{K_\pm}(x') r(xx') \overset{1}{K_\pm}(x) \nonumber \\
    \left\{ \overset{1}{K_\pm^T}(x^{-1}),\overset{2}{K_\pm^T}(x^{\prime-1}) \right\} = &\left[ r(x'/x), \overset{1}{K_\pm^T}(x^{-1}) \overset{2}{K_\pm^T}(x^{\prime-1}) \right] \\
    & + \overset{1}{K_\pm^T}(x^{-1}) r(x^{-1}x^{\prime-1}) \overset{2}{K_\pm^T}(x^{\prime-1}) - \overset{2}{K_\pm^T}(x^{\prime-1}) r(x^{-1}x^{\prime-1}) \overset{1}{K}_\pm(x^{-1}) \nonumber 
\end{align}
\end{subequations}
where $\overset{1}{K_\pm}(x)= K_\pm(x) \otimes I$ and $\overset{2}{K_\pm}(x) = I \otimes {K}_\pm$. 
The general solution to the reflection matrices are
\begin{align}
    K_+(x) & = \begin{pmatrix}
        \a_{+,1} x^{\frac12} + \a_{+,2} \g_+ x^{-\frac12} & \d_+ (x - \g_+ x^{-1}) - \b_{+,2} \\
        \b_{+,1} - x + \g_+ x^{-1} & \a_{+,2} x^{\frac12} + \a_{+,1} \g_+ x^{-\frac12}
    \end{pmatrix} , \\
    K_-(x) & = \begin{pmatrix}
        -\a_{-,1} x^{\frac12} - \a_{-,2} \g_- x^{-\frac12} & -x + \g_- x^{-1} + \b_{-,1} \\
        \b_{-,2} + \d_{-} (x - \g_- x^{-1}) & -\a_{-,2} x^{\frac12} - \a_{-,1}\g_- x^{-\frac12}
    \end{pmatrix}. 
\end{align}
with the parameters subjected to the constrains $\g_\pm^2=1$, $\d_{\pm}=0,1$ and:
\begin{align}\label{eq:K-constrain}
    \g_+\g_- = 1~,~ \b_{+,1} \left( 1 + \g_+ \right) = 0 = \b_{-,1}(1+\g_{-}) ~,~\b_{+,2} \left( 1 + \g_+ \right) = 0 = \b_{-,2}(1+\g_{-})~.
\end{align}
The transfer matrix takes the form 
\begin{align}
    \Tr~ T(x) = x^{N+2} + x^{N+2}+ \sum_{k=1}^{N+2} (-1)^k H_k (x^{N+2-k}+x^{k-N-2})
\end{align}
We also require the determinant of the reflection matrix to be coordinate-independent.
We would like to address that comparing to the type A case, the normalization of $\Tr~T(x)$ is trivial due to the fact that the contributions from $t_N(x)$ and $t_N^-(x)$ cancel each other, and the coefficient of $x$ in $K_\pm(x)$ are chosen as 1. 
The associated Hamiltonian is 
\begin{align}
    H_1 = H_A + J_+ + J_-~.
\end{align}
Here $H_A$ is the Haniltonian of the type $A$ open RTC
\begin{align}
    H_A = \sum_{n=1}^N 2\cosh p_n + 2\L^2 \sum_{n=1}^{N-1} e^{q_{n+1}-q_n} \cosh \frac{p_{n+1}+p_n}{2}~.
\end{align}
The contributions from the reflective boundaries are 
\begin{align}\label{eq:Hamiltonian-boundary}
\begin{split}
    J_+ & = \b_{+,1} + \a_{+,1} \L e^{-q_N-\frac{p_N}{2}} + \a_{+,2} \L e^{\frac{p_N}{2}-q_N} + \d_+ \L^2 e^{-2q_N}~,
    \\
    J_- & = \b_{-,1} + \a_{-,1} \L e^{q_1-\frac{p_1}{2}} + \a_{-,2} \L e^{\frac{p_1}{2}+q_1} + \d_- \L^2 e^{2q_1}~.
\end{split}
\end{align}

Let us first take the case $\g_+=\g_-=1$. 
Constraints in \eqref{eq:K-constrain} require all $\b_{\pm,1}=\b_{\pm,2}=0$.
Furthermore, $\a_{\pm,1}$ and $\a_{\pm,2}$ are constants, so that the determinants of the reflection matrices are coordinates independent. 
We scale the reflection matrices by an factor of $(x-x^{-1})^{-1}$:
\begin{align}\label{eq:K-prime}
\begin{split}
    K_+ & = \begin{pmatrix}
        \frac{\a_{+,1}+\a_{+,2}}{2(x^\frac12-x^{-\frac12})} - \frac{\a_{+,2}-\a_{+,1}}{2(x^\frac12+x^{-\frac12})} & \d_+ \\ -1 & \frac{\a_{+,2}+\a_{+,1}}{2(x^\frac12-x^{-\frac12})} + \frac{\a_{+,2}-\a_{+,1}}{2(x^\frac12+x^{-\frac12})}
    \end{pmatrix}~,~ \\
    K_- & = \begin{pmatrix}
        -\frac{\a_{-,1}+\a_{-,2}}{2(x^\frac12-x^{-\frac12})} + \frac{\a_{-,2}-\a_{-,1}}{2(x^\frac12+x^{-\frac12})} & -1 \\ \d_- & -\frac{\a_{-,2}+\a_{-,1}}{2(x^\frac12-x^{-\frac12})} - \frac{\a_{-,2}-\a_{-,1}}{2(x^\frac12+x^{-\frac12})}
    \end{pmatrix}~,~
\end{split}
\end{align}
so that the highest power of $x$ in the transfer matrix equals to the number of particles $N$. 

\begin{figure}[h!] \centering
\scalebox{1}{
\begin{tikzpicture}[square/.style={regular polygon,regular polygon sides=4},decoration={markings, 
    mark= at position 0.5 with {\arrow[scale=1.3]{stealth}}}
 ]
    
    \node at (-.5,-1) [style={circle,draw},thick,inner sep=4pt] (w1) {};
    \node at (.5,-1) [style={circle,draw},thick,inner sep=4pt] (w2) {};
    \node at (1.5,-1) [style={circle,draw},thick,inner sep=4pt] (w3) {};
    \node at (-1.5,-2) [style={circle,draw},thick,inner sep=4pt] (w4) {};
    \node at (-.5,-2) [style={circle,draw},thick,inner sep=4pt] (w5) {};
    \node at (.5,-2) [style={circle,draw},thick,inner sep=4pt] (w6) {};

    \draw[thick] (w1) -- (w2);
    \draw[thick,double] (w2) -- (w3);
    \draw[thick] (w1) -- (-1,-1);
    \draw[thick] (w6) -- (1,-2);
    \draw[thick,double] (w4) -- (w5);
    \draw[thick] (w5) -- (w6);

    \node at (-1,-2) {$\boldsymbol{>}$};
    \node at (1,-1) {$\boldsymbol{<}$};
    \node at (1.3,-2) {$\cdots$};
    \node at (-1.3,-1) {$\cdots$};

    \node[left] at (-2,-1) {$K_-^C$:};
    \node[left] at (-2,-2) {$K_+^C$:};
    
\end{tikzpicture}
}
\caption{Long root on the end of Dynkin diagram of Lie algebra $\fg$}\label{fig:long}
\end{figure}
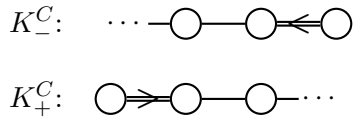

\paragraph{Type C boundary}
If the Lie algebra $\fg$ has long root at the end of the Dynkin diagram, we say $\fg$ has \emph{type C boundary}. See Fig.~\ref{fig:long}. 
The reflection matrix corresponds to type C boundary are
\begin{align}\label{eq:K-C}
\begin{split}
    K^C_+(x) & = {K_+(x,\a_{+,1}=\a_{+,2}=0,\d_+=1)} = \s~,~ \\
    K^C_-(x) & = {K_-(x,\a_{-,1}=\a_{-,2}=0,\d_-=1)} = \s^{-1}~.~
\end{split}
\end{align}
The contribution from the reflective boundary is 
\begin{align}
    J^C_+ = \L^2 e^{-2q_N}~,~ J^C_- = \L^2 e^{2q_1}~.
\end{align}

\begin{figure}[h!] \centering
\scalebox{1}{
\begin{tikzpicture}[square/.style={regular polygon,regular polygon sides=4},decoration={markings, 
    mark= at position 0.5 with {\arrow[scale=1.3]{stealth}}}
 ]
    
    \node at (-.5,-1) [style={circle,draw},thick,inner sep=4pt] (w1) {};
    \node at (.5,-1) [style={circle,draw},thick,inner sep=4pt] (w2) {};
    \node at (1.5,-1) [style={circle,draw},thick,inner sep=4pt] (w3) {};
    \node at (-1.5,-2) [style={circle,draw},thick,inner sep=4pt] (w4) {};
    \node at (-.5,-2) [style={circle,draw},thick,inner sep=4pt] (w5) {};
    \node at (.5,-2) [style={circle,draw},thick,inner sep=4pt] (w6) {};

    \draw[thick] (w1) -- (w2);
    \draw[thick,double] (w2) -- (w3);
    \draw[thick] (w1) -- (-1,-1);
    \draw[thick] (w6) -- (1,-2);
    \draw[thick,double] (w4) -- (w5);
    \draw[thick] (w5) -- (w6);

    \node at (-1,-2) {$\boldsymbol{<}$};
    \node at (1,-1) {$\boldsymbol{>}$};
    \node at (1.3,-2) {$\cdots$};
    \node at (-1.3,-1) {$\cdots$};

    \node[left] at (-2,-1) {$K_-^B$, $\bar{K}_-^B$:};
    \node[left] at (-2,-2) {$K_+^B$, $\bar{K}_-^B$:};
    
\end{tikzpicture}
}
\caption{Short roots on the end of Dynkin diagram of Lie algebra $\fg$}\label{fig:short}
\end{figure}
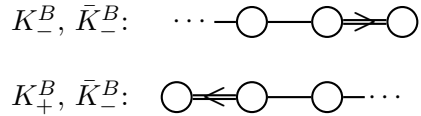

\paragraph{Type B boundary} We say the RTC has \emph{type B boundary} if the associated Lie algebra $\fg$ has short root at the end of the Dynkin diagram similar to $B_N$ like in Fig.~\ref{fig:short}. We consider two cases of the reflection matrices for a type B boundary. 
\paragraph{Type B-1:}
We take $\a_{\pm,1}=\a_{\pm,2}=\L$ and $\d_{\pm}=0$ in \eqref{eq:K-prime}.  
\begin{align}\label{eq:K-B1}
\begin{split}
    K^B_+(x) & = K_+(\a_{+,1}=\a_{+,2}=\L,\d_+=0) = \begin{pmatrix}
        \frac{\L}{x^{\frac12}-x^{-\frac12}} & 0 \\ -1 & \frac{\L}{x^{\frac12}-x^{-\frac12}}
    \end{pmatrix}~,~ \\
    K^B_-(x) & = K_-(\a_{-,1}=\a_{-,2}=\L,\d_-=0) = \begin{pmatrix}
        \frac{\L}{x^{\frac12}-x^{-\frac12}} & -1 \\ 0 & \frac{\L}{x^{\frac12}-x^{-\frac12}}
    \end{pmatrix}~.~ \\
\end{split}
\end{align}
The boundary contributions to the Hamiltonian are
\begin{align}
    J^B_+ = 2\L^2 e^{-q_N} \cosh \frac{p_N}{2} ~,~ J^B_- = 2\L^2 e^{q_1} \cosh \frac{p_1}{2} ~.~
\end{align} 
It is obvious that these boundary reflection matrices are obtained from the RTC Lax matrix \eqref{def:Lax} with frozen canonical variables.
\begin{align}
\begin{split}
    K^B_+(x) & = \frac{1}{x^{\frac12}-x^{-\frac12}} \s L^+(x,q=0,p=0)~,~ \\
    K_-^B(x) & = \frac{1}{x^{\frac12}-x^{-\frac12}} L^+(x,q=0,p=0) \s^{-1}~.
\end{split}
\end{align}

\paragraph{Type B-2:}
A more interesting case is when one of $\a_{\pm,1}$ and $\a_{\pm,2}$ vanishes and the other one equals to $\L$. Denote
\begin{align}
    \a_{\pm,2}+\a_{\pm,1} = \L~,~\a_{\pm,2}-\a_{\pm,1} = \k_\pm \L~,~ \k_\pm = \pm 1~,~\d_\pm=0~.
\end{align}
\begin{align}\label{eq:K-B2}
\begin{split}
    \bar{K}_+^B & = \frac{1}{x-x^{-1}} \begin{pmatrix}
        \L x^{-\frac{\k_+}{2}} & 0 \\ x^{-1}-x & \L x^{\frac{\k_+}{2}}
    \end{pmatrix} ~,~
    \bar{K}_-^B = \frac{1}{x-x^{-1}} \begin{pmatrix}
        -\L x^{-\frac{\k_-}{2}} & x^{-1}-x \\ 0 & -\L x^{\frac{\k_-}{2}}~
    \end{pmatrix}. 
\end{split}
\end{align}
The contribution of the reflective boundaries to the Hamiltonian is 
\begin{align}
    \bar{J}_+^B = \L^2 e^{-q_N+\k_+ \frac{p_N}{2}}~,~ \bar{J}^B_- = \L^2 e^{q_1+\k_- \frac{p_1}{2}}~.
\end{align}
In Section~\ref{sec:C-dual} we will see why this boundary is important. 

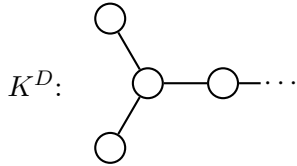
\begin{figure}[h!] \centering
\scalebox{1}{
\begin{tikzpicture}[square/.style={regular polygon,regular polygon sides=4},decoration={markings, 
    mark= at position 0.5 with {\arrow[scale=1.3]{stealth}}}
 ]
    
    \node at (.5,-2) [style={circle,draw},thick,inner sep=4pt] (w1) {};
    \node at (-.5,-2) [style={circle,draw},thick,inner sep=4pt] (w2) {};
    \node at (-1,-1.14) [style={circle,draw},thick,inner sep=4pt] (w3) {};
    \node at (-1,-2.86) [style={circle,draw},thick,inner sep=4pt] (w4) {};
    
    \draw[thick] (w1) -- (w2);
    \draw[thick] (w2) -- (w3);
    \draw[thick] (w2) -- (w4);
    \draw[thick] (w1) -- (1,-2);
    \node at (1.3,-2) {$\cdots$};

    \node at (-2,-2) {$K^D$:};
    
\end{tikzpicture}
}
\caption{Type D boundary of the Dynkin diagram}\label{fig:D-boundary}
\end{figure}

\paragraph{Type D boundary} Finally we consider the case where the reflection matrices are coordinates dependent $K_+(x,q_{N},p_{N})$, $K_-(x,q_1,p_1)$. This requires $\g_+=\g_-=-1$. 
If the reflection matrices are coordinate dependent, the type A RTC between the reflective boundaries are shortened so that the total number of particles in the RTC is fixed. 
\begin{align}
\begin{split}
    T(x) = & ~ K_+(x;q_N,p_N) t_{N-2}(x) K_-(x;q_1,p_1) t^-_{N-2}(x^{-1}). 
\end{split}
\end{align}
In particular, the known reflection matrices for type D boundaries \cite{Kuznetsov:1994ur,Lee:2024bqg} are:
\begin{align}\label{eq:K-D}
    & K^D (x;q,p) = \begin{pmatrix}
        x+x^{-1} - 2\cosh(p) & 2\L[\cosh(q+\frac{p}{2})x^{\frac12} - \cosh(q-\frac{p}{2}) x^{-\frac12}] \\
        2\L [\cosh(q+\frac{p}{2}) x^{-\frac12} -  \cosh(q-\frac{p}{2})x^{\frac12}] & \L^2[x + x^{\frac12} - 2\cosh(2q)]
    \end{pmatrix}  
\end{align}
with 
\begin{align}
    & K^D_+ (x;q,p) = \s K^D(x;q,p)~,~ 
     K^D_-(x;q,p) = K^D(x;q,p) \s^{-1}~.
\end{align}
Every type D boundary shortens the length of type A RTC between the reflective boundary by one so that the total number of particles in an RTC is fixed. The contributions from the reflective boundary to the Hamiltonian are as follows:
\begin{align}
\begin{split}
    {J}_+^D & = 2 \L^2 e^{-q_{N-1}-q_N}\cosh \frac{p_{N-1}-p_N}{2} + \L^4 e^{-2q_{N-1}} ;\\
    {J}_-^D & = 2\L^2 e^{q_1+q_2}\cosh \frac{p_1-p_2}{2} + \L^4 e^{2q_2}.
\end{split}
\end{align}
The root system associated to ${J}_\pm^D$ is of type D. See Fig.~\ref{fig:D-boundary}.  
Note that there are two $\L^4$ order term that are not originated from the fundamental root.

\section{Cluster integrable systems on dimer graphs}\label{sec:dimer}

RTC arises naturally from the Lie algebra and therefore has a cluster description \cite{Kenyon2011dimers,Marshakov:2012kv}. 
In this section, we review the cluster integrable system associated to a dimer graph on a torus. 

A convex polygon $\Delta$ with vertexes in $\BZ^2 \subset \BR^2$ can be considered as the Newton polygon of the polynomial $f_\Delta(x,y)$, and
\begin{align}
    f_\Delta(x,y) = \sum_{(a,b)\in \Delta} x^a y^b f_{a,b} = 0
\end{align}
defines a spectral curve in $\BC^\times \times \BC^\times$. The genus of the curve equals to the number of points strictly inside the polygon $\Delta$. 

A convex Newton polygon $\Delta$, modulo action of $SA(2,\BZ) = SL(2,\BZ) \ltimes \BZ_2$, defines an integrable system on X-cluster Poisson variety $\CalX$ of dimension $2S$, where $S$ is the area of the polygon $\Delta$. The Poisson structure is encoded in a quiver $\CalQ$ with $2S$ vertices. 


A cluster algebra is defined by a \emph{cluster seed} $\Sigma$. A seed is a triplet $\Sigma=(I,I_0,\ve)$, where $I$ is a finite set, $I_0 \subset I$ is a subset, and $\ve=(\ve_{i,j})_{i,j\in I}$ is a skew-symmetric $\BZ$-valued matrix such that $\ve_{i,j} \in \BZ$ unless $i,j \in I_0$. $\ve_{i,j}$ is the number of arrows from the $i$-th to the $j$-th vertex in $\CalQ$. 
To a given seed $\Sigma$, we associated an algebraic torus $(\BC^\times)^{|I|}$, called $\CalX$-cluster torus. Its coordinates $(f_i)_{i\in I}$ are called the cluster variables. The logarithmically constant Poisson bracket takes the form 
\begin{align}
    \{f_i,f_j\} = \ve_{i,j} f_if_j~.
\end{align}

The graph dual of the quiver $\CalQ$ is a bipartite graph on a torus, known as \emph{dimer model} $\G$. 
The cluster variables are represented by clockwise loops surrounding the corresponding faces on the dimer graph, which we also call \emph{face variables}.

\paragraph{Examples of Newton Polygon} In this note the RTCs correspond to the following Newton polygon type: 
\begin{figure}[h!] \centering
\scalebox{0.7}{
\begin{tikzpicture}[square/.style={regular polygon,regular polygon sides=4},decoration={markings, 
    mark= at position 0.5 with {\arrow[scale=1.3]{stealth}}}
 ]
     \node at (-4,-2) [circle,fill,inner sep=1pt] (b21) {};
     \node at (-3,-2) [circle,fill,inner sep=1pt] (b22) {};
     \node at (-2,-2) [circle,fill,inner sep=1pt] (b23) {};
     \node at (-1,-2) [circle,fill,inner sep=1pt] (b24) {};
     \node at (1,-2) [circle,fill,inner sep=1pt] (b25) {};
     \node at (2,-2) [circle,fill,inner sep=1pt] (b26) {};
     \node at (3,-2) [circle,fill,inner sep=1pt] (b27) {};
     \node at (4,-2) [circle,fill,inner sep=1pt] (b28) {};

     \node at (-4,-1) [circle,fill,inner sep=1pt] (b11) {};
     \node at (-3,-1) [circle,fill,inner sep=1pt] (b12) {};
     \node at (-2,-1) [circle,fill,inner sep=1pt] (b13) {};
     \node at (-1,-1) [circle,fill,inner sep=1pt] (b14) {};
     \node at (1,-1) [circle,fill,inner sep=1pt] (b15) {};
     \node at (2,-1) [circle,fill,inner sep=1pt] (b16) {};
     \node at (3,-1) [circle,fill,inner sep=1pt] (b17) {};
     \node at (4,-1) [circle,fill,inner sep=1pt] (b18) {};

     \node at (-4,-3) [circle,fill,inner sep=1pt] (b31) {};
     \node at (-3,-3) [circle,fill,inner sep=1pt] (b32) {};
     \node at (-2,-3) [circle,fill,inner sep=1pt] (b33) {};
     \node at (-1,-3) [circle,fill,inner sep=1pt] (b34) {};
     \node at (1,-3) [circle,fill,inner sep=1pt] (b35) {};
     \node at (2,-3) [circle,fill,inner sep=1pt] (b36) {};
     \node at (3,-3) [circle,fill,inner sep=1pt] (b37) {};
     \node at (4,-3) [circle,fill,inner sep=1pt] (b38) {};
     
     \node at (-4,-4) {$|$};
     \node at (4,-4) {$|$};
    \draw[thick] (-4,-4) -- (4,-4);
     \node at (0,-1) {$\cdots$};
     \node at (0,-2) {$\cdots$};
     \node at (0,-3) {$\cdots$};
    \node at (0,-3.8) {$N$};

    \draw[thick] (b21) -- (b11);
    \draw[thick] (b11) -- (b28);
    \draw[thick] (b28) -- (b38);
    \draw[thick] (b38) -- (b21);

    \node at (-5,-2) {$Y^{N,0}$};

\end{tikzpicture}
}
\caption{The Newton polygon for Toda chain on $N$ sites.}\label{fig:Newton polygon}
\end{figure}
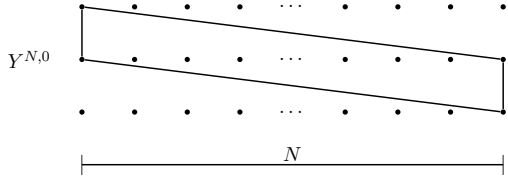

Different dimer graphs can give rise to the same cluster integrable system, in the same spirit that the 5-brane web engineering of a 5d SCFT is not unique. The $SL(2,\BZ)$ duality of type IIB String Theory is equivalent to the $SL(2,\BZ) \subset SA(2,\BZ)$ action on the cluster integrable system. On the level of the spectral curve it corresponds to the following transformation of the spectral parameters:
\begin{align}\label{eq:sl2z-trans}
	(x,y) \mapsto (x^ay^b,x^c y^d) ~,~ \begin{pmatrix} a & b \\ c & d \end{pmatrix} \in SL(2,\BZ)
\end{align}
More interestingly, the Hanany-Witten move \cite{Hanany:1996ie,Bergman:1999na} in type IIB String Theory is equivalent to the birational transformation of cluster algebra \cite{Kho:2025fmp,Kho:2025jxk} by taking
\begin{align}\label{eq:bira-trans}
	(x,y) \mapsto (x,(x+a)y)~
\end{align}
for the spectral parameters. 
In section. \ref{sec:example} we will use this fact frequently by establishing equivalence between spectral curves obtained from Sklyanin's Lax matrix formalism and the dimer model Kasteleyn matrix through both the $SL(2,\BZ)$ action and birational transformation. 

The third equivalence between the cluster integrable system on a dimer is established through cluster mutation. 
For given seeds $\Sigma=(I,I_0,\ve)$ and $\Sigma'=(I',I_0',\ve')$, fix $k \in I\backslash I_0$. An isomorphism $\mu_k:\Sigma \to \Sigma'$ is called \emph{seed mutation in direction $k$} if $\mu_k(I_0) = I_0'$ and
\begin{align}
    \ve'_{ij} = \begin{cases}
        -\ve_{ij} & \text{if $i=k$ or $j=k$} \\
        \ve_{ij} & \ve_{ik}\ve_{kj} \leq 0\\
        \ve_{ij} + |\ve_{ik}| \ve_{kj} & \ve_{ik}\ve_{kj}>0
    \end{cases}.
\end{align}
For a seed mutation $\mu_k$, we define cluster mutation $\mu_k^c: \CalX_\Sigma \to \CalX_{\Sigma'}$ by
\begin{align}
    \mu_k^c ( f_i) = \begin{cases}
        f_i^{-1} & \text{if } i=k \\
        f_i (1+f_k)^{\ve_{ik}} & \text{if } i\neq k \text{ and } \ve_{ik}\geq 0 \\
        f_i (1+f_k^{-1})^{\ve_{ik}} & \text{if } i\neq k \text{ and } \ve_{ik}\leq 0
    \end{cases}
\end{align}

A cluster algebra associated to a seed $\S$ is defined as the subalgebra of the algebra $\CalX$ consisting of universally Laurent elements, i.e., the ones that remain Laurent polynomials under all finite sequences of cluster mutations.

The cluster integrable systems defined from the convex Newton polygon $\Delta$ are invariant under seed mutation, despite the dimer $\G$ and quiver $\CalQ$ are different. 

\subsection{Type A RTC}
\begin{figure}[h!] \centering
\scalebox{0.8}{
\begin{tikzpicture}[square/.style={regular polygon,regular polygon sides=4},decoration={markings, 
    mark= at position 0.5 with {\arrow[scale=1.3]{stealth}}}
 ]
     \node at (-4,-2) [circle,fill,inner sep=3pt] (b1) {};
     \node at (-3.7,-1.7) {${}_{1}$};
      \node at (-4,-4) [circle,fill,inner sep=3pt] (b2) {};
     \node at (-3.7,-3.7) {${}_{2}$} ;
     \node at (-4,-6) [circle,fill,inner sep=3pt] (b3) {};
     \node at (-3.7,-5.7) {${}_{3}$} ;
     \node at (-4,-10) [circle,fill,inner sep=3pt] (b4) {};
    \node at (-3.7,-9.7) {${}_{N-1}$} ;
    \node at (-4,-12) [circle,fill,inner sep=3pt] (b5) {};
    \node at (-3.7,-11.7) {${}_{N}$} ;

     \node at (2,-2) [circle,fill,inner sep=3pt] (b6) {};
     \node at (2.3,-1.7) {${}_{1}$};
      \node at (2,-4) [circle,fill,inner sep=3pt] (b7) {};
     \node at (2.3,-3.7) {${}_{2}$} ;
     \node at (2,-6) [circle,fill,inner sep=3pt] (b8) {};
     \node at (2.3,-5.7) {${}_{3}$} ;
     \node at (2,-10) [circle,fill,inner sep=3pt] (b9) {};
    \node at (2.3,-9.7) {${}_{N-1}$} ;
    \node at (2,-12) [circle,fill,inner sep=3pt] (b10) {};
    \node at (2.3,-11.7) {${}_{N}$} ;

    \node at (-2,-2) [style={circle,draw},thick,inner sep=3pt] (w1) {};
    \node at (-1.7,-1.7) {${}_{1}$} ;
    \node at (-2,-4) [style={circle,draw},thick,inner sep=3pt] (w2) {};
    \node at (-1.7,-3.7) {${}_{2}$} ;
     \node at (-2,-6) [style={circle,draw},thick,inner sep=3pt] (w3) {};
     \node at (-1.7,-5.7) {${}_{3}$} ;
     \node at (-2,-10) [style={circle,draw},thick,inner sep=3pt] (w4) {};
     \node at (-1.7,-9.7) {${}_{N-1}$} ;
     \node at (-2,-12) [style={circle,draw},thick,inner sep=3pt] (w5) {};
     \node at (-1.7,-11.7) {${}_{N}$} ;

    \node at (4,-2) [style={circle,draw},thick,inner sep=3pt] (w6) {};
    \node at (4.3,-1.7) {${}_{1}$} ;
    \node at (4,-4) [style={circle,draw},thick,inner sep=3pt] (w7) {};
    \node at (4.3,-3.7) {${}_{2}$} ;
     \node at (4,-6) [style={circle,draw},thick,inner sep=3pt] (w8) {};
     \node at (4.3,-5.7) {${}_{3}$} ;
     \node at (4,-10) [style={circle,draw},thick,inner sep=3pt] (w9) {};
     \node at (4.3,-9.7) {${}_{N-1}$} ;
     \node at (4,-12) [style={circle,draw},thick,inner sep=3pt] (w10) {};
     \node at (4.3,-11.7) {${}_{N}$} ;

     \node at (-3,-8) {$\vdots$} ;
     \node at (3,-8) {$\vdots$} ;

    \draw[postaction={decorate},thick] (w1)--(b1);
    \draw[postaction={decorate},thick] (w2)--(b2);
    \draw[postaction={decorate},thick] (w3)--(b3);
    \draw[postaction={decorate},thick] (w4)--(b4);
    \draw[postaction={decorate},thick] (w5)--(b5);

    \draw[postaction={decorate},thick] (w1)--(-1,-2);
    \draw[postaction={decorate},thick] (w2)--(-1,-4);
    \draw[postaction={decorate},thick] (w3)--(-1,-6);
    \draw[postaction={decorate},thick] (w4)--(-1,-10);
    \draw[postaction={decorate},thick] (w5)--(-1,-12);

    \draw[postaction={decorate},thick] (-5,-2)--(b1);
    \draw[postaction={decorate},thick] (-5,-4)--(b2);
    \draw[postaction={decorate},thick] (-5,-6)--(b3);
    \draw[postaction={decorate},thick] (-5,-10)--(b4);
    \draw[postaction={decorate},thick] (-5,-12)--(b5);

    \draw[postaction={decorate},thick] (w2)--(b1);
    \draw[postaction={decorate},thick] (w3)--(b2);
    \draw[postaction={decorate},thick] (w5)--(b4);
    \draw[postaction={decorate},thick] (w1)--(-3,-1);
    \draw[postaction={decorate},thick] (w4)--(-3,-9);

    \draw[postaction={decorate},thick] (w1)--(-1,-3);
    \draw[postaction={decorate},thick] (w2)--(-1,-5);
    \draw[postaction={decorate},thick] (w3)--(-1,-7);
    \draw[postaction={decorate},thick] (w4)--(-1,-11);
    \draw[postaction={decorate},thick] (w5)--(-1,-13);
    
    \draw[postaction={decorate},thick] (-5,-1)--(b1);
    \draw[postaction={decorate},thick] (-5,-3)--(b2);
    \draw[postaction={decorate},thick] (-5,-5)--(b3);
    \draw[postaction={decorate},thick] (-5,-9)--(b4);
    \draw[postaction={decorate},thick] (-5,-11)--(b5);

    \draw[postaction={decorate},thick] (-3,-7)--(b3);
    \draw[postaction={decorate},thick] (-3,-13)--(b5); 

    \draw[postaction={decorate},thick] (w6)--(b6);
    \draw[postaction={decorate},thick] (w7)--(b7);
    \draw[postaction={decorate},thick] (w8)--(b8);
    \draw[postaction={decorate},thick] (w9)--(b9);
    \draw[postaction={decorate},thick] (w10)--(b10);

    \draw[postaction={decorate},thick] (w6)--(5,-2);
    \draw[postaction={decorate},thick] (w7)--(5,-4);
    \draw[postaction={decorate},thick] (w8)--(5,-6);
    \draw[postaction={decorate},thick] (w9)--(5,-10);
    \draw[postaction={decorate},thick] (w10)--(5,-12);

    \draw[postaction={decorate},thick] (1,-2)--(b6);
    \draw[postaction={decorate},thick] (1,-4)--(b7);
    \draw[postaction={decorate},thick] (1,-6)--(b8);
    \draw[postaction={decorate},thick] (1,-10)--(b9);
    \draw[postaction={decorate},thick] (1,-12)--(b10);

    \draw[postaction={decorate},thick] (w7)--(b6);
    \draw[postaction={decorate},thick] (w8)--(b7);
    \draw[postaction={decorate},thick] (w10)--(b9);
    \draw[postaction={decorate},thick] (w6)--(3,-1);
    \draw[postaction={decorate},thick] (w9)--(3,-9);

    \draw[postaction={decorate},thick] (w6)--(5,-3);
    \draw[postaction={decorate},thick] (w7)--(5,-5);
    \draw[postaction={decorate},thick] (w8)--(5,-7);
    \draw[postaction={decorate},thick] (w9)--(5,-11);
    \draw[postaction={decorate},thick] (w10)--(5,-13);
    
    \draw[postaction={decorate},thick] (1,-1)--(b6);
    \draw[postaction={decorate},thick] (1,-3)--(b7);
    \draw[postaction={decorate},thick] (1,-5)--(b8);
    \draw[postaction={decorate},thick] (1,-9)--(b9);
    \draw[postaction={decorate},thick] (1,-11)--(b10);

    \draw[postaction={decorate},thick] (3,-7)--(b8);
    \draw[postaction={decorate},thick] (3,-13)--(b10);

          



    \node (t0) at (-5,-1) {};
    \node (t3) at (5,-1) {};
    \node (t4) at (-2,-1) {};
    \node (t5) at (2,-1) {};
     \node (t1) at (-1,-1) {};
     \node (t2) at (1,-1) {};

     \node (bb1) at (-1,-13) {};
         \node (bb2) at (1,-13) {};
          \node (bb3) at (-2,-13) {};
     \node (bb4) at (2,-13) {};
          \node (bb5) at (-5,-13) {};
     \node (bb6) at (5,-13) {};

     \draw[dotted,very thick,red] (t0) -- (t1);
        \draw[dotted,very thick,red] (bb1) -- (bb5);
       \draw[dotted,very thick,blue] (t0) -- (bb5);
        \draw[dotted,very thick,blue] (t1) -- (bb1);
        
        \draw[dotted,very thick,red] (bb2) -- (bb6);
        \draw[dotted,very thick,red] (t2) -- (t3);
       \draw[dotted,very thick,blue] (t2) -- (bb2);
        \draw[dotted,very thick,blue] (t3) -- (bb6);

    \node at (-2,-3) {$f_1^\times$};
    \node at (-4,-3) {$f_1^+$};
    \node at (-2,-5) {$f_2^\times$};
    \node at (-4,-5) {$f_2^+$};

    \node at (-2,-7) {$f_3^\times$};
    \node at (-4,-7) {$f_3^+$};

    \node at (-2,-9) {$f_{N-2}^\times$};
    \node at (-4,-9) {$f_{N-2}^+$};
    \node at (-2,-11) {$f_{N-1}^\times$};
    \node at (-4,-11) {$f_{N-1}^+$};

    \node at (-2,-1) {$f_N^\times$};
    \node at (-4,-1) {$f_N^+$};

    \node at (-2,-13) {$f_N^\times$};
    \node at (-4,-13) {$f_N^+$};


    \node at (3,-.7) {$\L\xi_1$};
    \node[left] at (3,-3) {$\L\xi_2$};
    \node[left] at (3,-5) {$\L\xi_3$};
    \node at (3,-7.2) {$\L\xi_4$};
    \node at (3,-8.7) {$\L\xi_{N-1}$};
    \node[left] at (3,-11) {$\L\xi_N$};

    \node at (3.3,-2.3) {$\eta_1^{-1}$};
    \node at (3.3,-4.3) {$\eta_2^{-1}$};
    \node at (3.3,-6.3) {$\eta_3^{-1}$};
    \node at (3.3,-10.3) {$\eta_{N-1}^{-1}$};
    \node at (3.3,-12.3) {$\eta_N^{-1}$};

    \node[right] at (5,-2) {${\eta_1}$};
    \node[right] at (5,-4) {${\eta_2}$};
    \node[right] at (5,-6) {${\eta_3}$};
    \node[right] at (5,-10) {${\eta_{N-1}}$};
    \node[right] at (5,-12) {${\eta_N}$};

    \node[left] at (1,-2) {${\eta_1}$};
    \node[left] at (1,-4) {${\eta_2}$};
    \node[left] at (1,-6) {${\eta_3}$};
    \node[left] at (1,-10) {${\eta_{N-1}}$};
    \node[left] at (1,-12) {${\eta_N}$};

    \node[right] at (5,-3) {${\L\xi^{-1}_1}$};
    \node[right] at (5,-5) {${\L\xi^{-1}_2}$};
    \node[right] at (5,-7.2) {${\L\xi^{-1}_3}$};
    \node[right] at (5,-11) {${\L\xi^{-1}_{N-1}}$};
    \node[right] at (5,-13.2) {${\L\xi^{-1}_N}$};

    \node[left] at (1,-1) {${\L\xi^{-1}_N}$};
    \node[left] at (1,-3) {${\L\xi^{-1}_1}$};
    \node[left] at (1,-5) {${\L\xi^{-1}_2}$};
    \node[left] at (1,-8.8) {${\L\xi^{-1}_{N-2}}$};
    \node[left] at (1,-11) {${\L\xi^{-1}_{N-1}}$};

    \node[right] at (3,-13.2) {$\L\xi_1$};

\end{tikzpicture}
}
\caption{The bipartite graph $Y^{N,0}$ model associated with type A RTC with Lie algebra ${A}_N^{(1)}$. The horizontal dotted lines (red) on the top and the bottom are identified, as well as the vertical dotted lines (blue) on the left and right. In turn, the bipartite graph is drawn on a torus $T^2$. \\
Left: face variables. Right: gauged edge variables. }\label{fig:bipartite}
\end{figure}
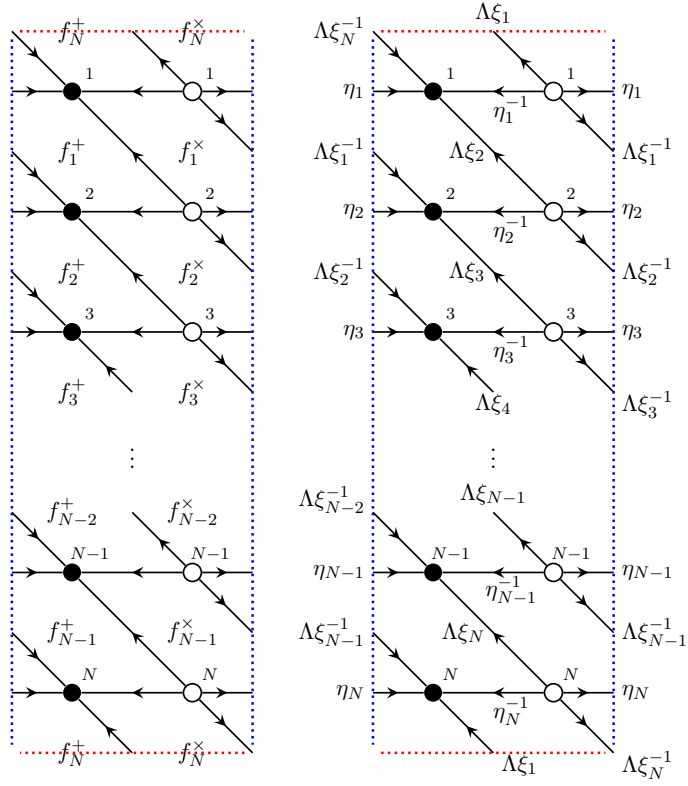

A typical bipartite dimer graph for affine ${A}_N^{(1)}$ (aka $\hat{A}_N$) \footnote{In this note we will mostly use Victor Kac's labeling for the (twisted) Lie algebra \cite{kac1990infinite}.} RTC is shown at Fig.~\ref{fig:bipartite}. The dimer graph is also known as the $Y^{N,0}$ model. For a Toda system of $N$ particles the dimer graph has $2N$ faces , $2N$ vertices $\{w_n\}_{n=1}^N$, $\{b_n\}_{n=1}^N$, and $4N$ edges. 

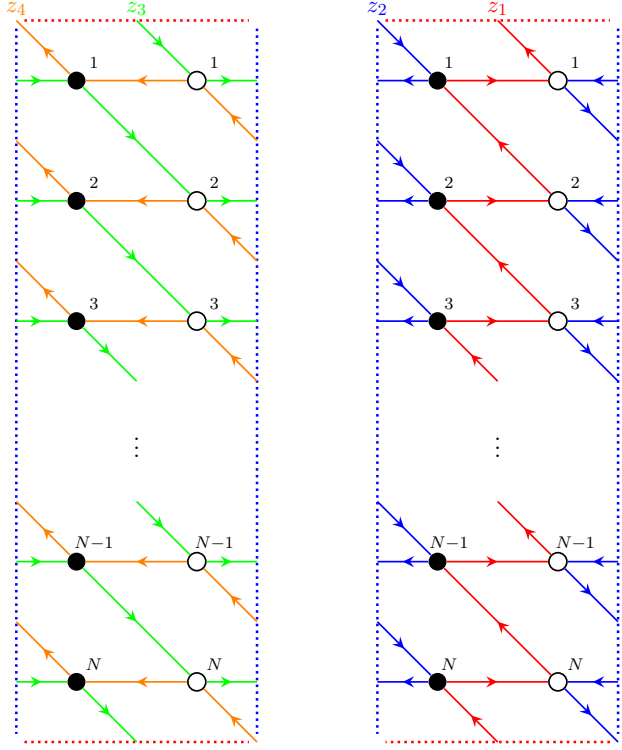
\begin{figure}[h!] \centering
\scalebox{0.8}{
\begin{tikzpicture}[square/.style={regular polygon,regular polygon sides=4},decoration={markings, 
    mark= at position 0.5 with {\arrow[scale=1.3]{stealth}}}
 ]

     \node at (2,-2) [circle,fill,inner sep=3pt] (b6) {};
     \node at (2.3,-1.7) {${}_{1}$};
      \node at (2,-4) [circle,fill,inner sep=3pt] (b7) {};
     \node at (2.3,-3.7) {${}_{2}$} ;
     \node at (2,-6) [circle,fill,inner sep=3pt] (b8) {};
     \node at (2.3,-5.7) {${}_{3}$} ;
     \node at (2,-10) [circle,fill,inner sep=3pt] (b9) {};
    \node at (2.3,-9.7) {${}_{N-1}$} ;
    \node at (2,-12) [circle,fill,inner sep=3pt] (b10) {};
    \node at (2.3,-11.7) {${}_{N}$} ;

     \node at (8,-2) [circle,fill,inner sep=3pt] (b11) {};
     \node at (8.2,-1.7) {${}_{1}$};
      \node at (8,-4) [circle,fill,inner sep=3pt] (b12) {};
     \node at (8.2,-3.7) {${}_{2}$} ;
     \node at (8,-6) [circle,fill,inner sep=3pt] (b13) {};
     \node at (8.2,-5.7) {${}_{3}$} ;
     \node at (8,-10) [circle,fill,inner sep=3pt] (b14) {};
    \node at (8.2,-9.7) {${}_{N-1}$} ;
    \node at (8,-12) [circle,fill,inner sep=3pt] (b15) {};
    \node at (8.2,-11.7) {${}_{N}$} ;

    \node at (4,-2) [style={circle,draw},thick,inner sep=3pt] (w6) {};
    \node at (4.3,-1.7) {${}_{1}$} ;
    \node at (4,-4) [style={circle,draw},thick,inner sep=3pt] (w7) {};
    \node at (4.3,-3.7) {${}_{2}$} ;
     \node at (4,-6) [style={circle,draw},thick,inner sep=3pt] (w8) {};
     \node at (4.3,-5.7) {${}_{3}$} ;
     \node at (4,-10) [style={circle,draw},thick,inner sep=3pt] (w9) {};
     \node at (4.3,-9.7) {${}_{N-1}$} ;
     \node at (4,-12) [style={circle,draw},thick,inner sep=3pt] (w10) {};
     \node at (4.3,-11.7) {${}_{N}$} ;

    \node at (10,-2) [style={circle,draw},thick,inner sep=3pt] (w11) {};
    \node at (10.3,-1.7) {${}_{1}$} ;
    \node at (10,-4) [style={circle,draw},thick,inner sep=3pt] (w12) {};
    \node at (10.3,-3.7) {${}_{2}$} ;
     \node at (10,-6) [style={circle,draw},thick,inner sep=3pt] (w13) {};
     \node at (10.3,-5.7) {${}_{3}$} ;
     \node at (10,-10) [style={circle,draw},thick,inner sep=3pt] (w14) {};
     \node at (10.3,-9.7) {${}_{N-1}$} ;
     \node at (10,-12) [style={circle,draw},thick,inner sep=3pt] (w15) {};
     \node at (10.3,-11.7) {${}_{N}$} ;

     \node at (3,-8) {$\vdots$} ;
     \node at (9,-8) {$\vdots$} ;

          



    \node (t0) at (7,-1) {};
    \node (t3) at (5,-1) {};
    \node (t5) at (2,-1) {};
     \node (t1) at (11,-1) {};
     \node (t2) at (1,-1) {};

     \node (bb1) at (11,-13) {};
         \node (bb2) at (1,-13) {};
     \node (bb4) at (2,-13) {};
          \node (bb5) at (7,-13) {};
     \node (bb6) at (5,-13) {};

     \draw[dotted,very thick,red] (t0) -- (t1);
        \draw[dotted,very thick,red] (bb1) -- (bb5);
       \draw[dotted,very thick,blue] (t0) -- (bb5);
        \draw[dotted,very thick,blue] (t1) -- (bb1);
        
        \draw[dotted,very thick,red] (bb2) -- (bb6);
        \draw[dotted,very thick,red] (t2) -- (t3);
       \draw[dotted,very thick,blue] (t2) -- (bb2);
        \draw[dotted,very thick,blue] (t3) -- (bb6);


    \draw[postaction={decorate},thick,red] (9,-13) -- (b15);
    \draw[postaction={decorate},thick,red] (b15) -- (w15);
    \draw[postaction={decorate},thick,red] (w15) -- (b14);
    \draw[postaction={decorate},thick,red] (b14) -- (w14);
    \draw[postaction={decorate},thick,red] (w14) -- (9,-9);
    \draw[postaction={decorate},thick,red] (9,-7) -- (b13);
    \draw[postaction={decorate},thick,red] (b13) -- (w13);
    \draw[postaction={decorate},thick,red] (w13) -- (b12);
    \draw[postaction={decorate},thick,red] (b12) -- (w12);
    \draw[postaction={decorate},thick,red] (w12) -- (b11);
    \draw[postaction={decorate},thick,red] (b11) -- (w11);
    \draw[postaction={decorate},thick,red] (w11) -- (9,-1);

    \draw[postaction={decorate},thick,blue] (7,-1) -- (b11);
    \draw[postaction={decorate},thick,blue] (b11) -- (7,-2);
    \draw[postaction={decorate},thick,blue] (7,-3) -- (b12);
    \draw[postaction={decorate},thick,blue] (b12) -- (7,-4);
    \draw[postaction={decorate},thick,blue] (7,-5) -- (b13);
    \draw[postaction={decorate},thick,blue] (b13) -- (7,-6);
    \draw[postaction={decorate},thick,blue] (7,-9) -- (b14);
    \draw[postaction={decorate},thick,blue] (b14) -- (7,-10);
    \draw[postaction={decorate},thick,blue] (7,-11) -- (b15);
    \draw[postaction={decorate},thick,blue] (b15) -- (7,-12);

    \draw[postaction={decorate},thick,blue] (11,-2) -- (w11);
    \draw[postaction={decorate},thick,blue] (w11) -- (11,-3);
    \draw[postaction={decorate},thick,blue] (11,-4) -- (w12);
    \draw[postaction={decorate},thick,blue] (w12) -- (11,-5);
    \draw[postaction={decorate},thick,blue] (11,-6) -- (w13);
    \draw[postaction={decorate},thick,blue] (w13) -- (11,-7);
    \draw[postaction={decorate},thick,blue] (11,-10) -- (w14);
    \draw[postaction={decorate},thick,blue] (w14) -- (11,-11);
    \draw[postaction={decorate},thick,blue] (11,-12) -- (w15);
    \draw[postaction={decorate},thick,blue] (w15) -- (11,-13);

    \node at (9,-.8) {\color{red} $z_1$};
    \node at (7,-.8) {\color{blue} $z_2$};


    \draw[postaction={decorate},thick,orange] (5,-13) -- (w10);
    \draw[postaction={decorate},thick,orange] (w10) -- (b10);
    \draw[postaction={decorate},thick,orange] (b10) -- (1,-11);
    \draw[postaction={decorate},thick,orange] (5,-11) -- (w9);
    \draw[postaction={decorate},thick,orange] (w9) -- (b9);
    \draw[postaction={decorate},thick,orange] (b9) -- (1,-9);
    \draw[postaction={decorate},thick,orange] (5,-7) -- (w8);
    \draw[postaction={decorate},thick,orange] (w8) -- (b8);
    \draw[postaction={decorate},thick,orange] (b8) -- (1,-5);
    \draw[postaction={decorate},thick,orange] (5,-5) -- (w7);
    \draw[postaction={decorate},thick,orange] (w7) -- (b7);
    \draw[postaction={decorate},thick,orange] (b7) -- (1,-3);
    \draw[postaction={decorate},thick,orange] (5,-3) -- (w6);
    \draw[postaction={decorate},thick,orange] (w6) -- (b6);
    \draw[postaction={decorate},thick,orange] (b6) -- (1,-1);

    \draw[postaction={decorate},thick,green] (3,-1) -- (w6);
    \draw[postaction={decorate},thick,green] (w6) -- (5,-2);
    \draw[postaction={decorate},thick,green] (1,-2) -- (b6);
    \draw[postaction={decorate},thick,green] (b6) -- (w7);
    \draw[postaction={decorate},thick,green] (w7) -- (5,-4);
    \draw[postaction={decorate},thick,green] (1,-4) -- (b7);
    \draw[postaction={decorate},thick,green] (b7) -- (w8);
    \draw[postaction={decorate},thick,green] (w8) -- (5,-6);
    \draw[postaction={decorate},thick,green] (1,-6) -- (b8);
    \draw[postaction={decorate},thick,green] (b8) -- (3,-7);
    \draw[postaction={decorate},thick,green] (3,-9) -- (w9);
    \draw[postaction={decorate},thick,green] (w9) -- (5,-10);
    \draw[postaction={decorate},thick,green] (1,-10) -- (b9);
    \draw[postaction={decorate},thick,green] (b9) -- (w10);
    \draw[postaction={decorate},thick,green] (w10) -- (5,-12);
    \draw[postaction={decorate},thick,green] (1,-12) -- (b10);
    \draw[postaction={decorate},thick,green] (b10) -- (3,-13);

    \node at (3,-.8) {\color{green} $z_3$};
    \node at (1,-.8) {\color{orange} $z_4$};
    
\end{tikzpicture}
}
\caption{The 4 zigzag loops of $Y^{N,0}$ dimer graph }\label{fig:A-zigzag}
\end{figure}

The cluster Poisson brackets for the dimer face variables is 
\begin{align}
    \{f_n^+,f_m^+\} = \{f_n^\times,f_m^\times\} = 0~,~ \{f_n^\times,f_m^+\} = (\d_{n,m+1}+\d_{n,m-1}-2\d_{n,m}) f_n^\times f_m^+~, ~n,m \in\BZ_N.
\end{align}
where in the non-vanishing r.h.s one can immediately recognize the Cartan matrix of $\hat{\mathfrak{sl}}_N$. 
This Poisson bracket has two obvious Casimir functions, which we choose as
\begin{align}
    \prod_{n=1}^N (f_n^\times f^+_n) = 1~,~ \prod_{n=1}^N f_n^\times = \L^{2N} = {z_1z_2}.
\end{align}
$z_{1,2}$ are zigzag loops (see Fig.~\ref{fig:A-zigzag}, right). The zigzag loops are paths that turn right most at black nodes and turn left most at white nodes. 
The total number of zigzag loops on a dimer graph $\G(\D)$ equals to the number of external vertices of the Newton polygon $\D$. There are four zigzag loops $z_{1,2,3,4}$ for $Y^{N,0}$ model. See  Fig.~\ref{fig:A-zigzag} for illustration.
The zigzag loops belong to the center of the cluster algebra $\CalX$, i.e. 
\begin{align}
    \{z_{1,2,3,4},f^+_i\} = 0 = \{z_{1,2,3,4},f^\times_i\} ~.
\end{align}
There is a single Casimir - diagonal twist of the monodromy matrix or coupling of the affine Toda chain. Reduction from the 4 zigzag loops to a single Casimir is a reminiscence of the freedom scaling $y \to ay$, $x \to bx$ and the fact that $z_1z_2z_3z_4=1$.

For the remainder of this note, we will mostly work with the edge variables, which do not have a canonical Poisson bracket. 
We choose the default orientation of the edge variables from white to black. 
Following \cite{Marshakov:2012kv,Jeong:2025yys}, we fix a gauge and express all edge variables by exponentiated Darboux coordinates $\xi_n$ and $\eta_n$: 
\begin{align}
    \{\xi_n,\eta_m\} = \frac12 \d_{n,m} \xi_n \eta_m~
\end{align}
so that the face variables are expressed, in terms of the oriented edge variables, as
\begin{align}
    f^\times_n = \L^2 \xi_{n}\xi_{n+1}^{-1}\eta_{n+1}\eta_n^{-1} ~,~ f_n^+ = \L^{-2} \xi_{n+1}\xi_n^{-1} \eta_n^{-1}\eta_{n+1}. 
\end{align}

The Kasteleyn matrix, a weighted adjancy matrix of the dimer graph, is given by an $N \times N$ matrix: 
\begin{align}
    \fD = \sum_{n=1}^N (\eta_n x^{-1}-\eta^{-1}_n) E_{n,n} + \L\xi_n E_{n,n-1} + \L \xi^{-1}_n x^{-1} E_{n,n+1}
\end{align}
$E_{m,n}$ is the $N\times N$ matrix with 1 at position $(m,n)$ and 0 elsewhere. 
We have additionally defined 
\begin{align}
    E_{N,N+1} = y^{-1} E_{N,1}~,~ E_{1,0} = y E_{1,N}~.~
\end{align}
The Kasteleyn matrix almost coincide with the $N \times N$ Lax matrix formalism of the type A RTC \cite{Ruijsenaars_RToda1990,Kuznetsov:1994ur,Lee:2023wbf}. The spectral curve of type A RTC is given by
\begin{align}\label{eq:spectral-A}
    0 = \det \fD (x,y) = -\frac{\L^Ny}{x^N} \prod_{n=1}^N \xi_n - \frac{\L^N}{y} \prod_{n=1}^{N} \xi_n^{-1} + (-1)^N \prod_{n=1}^N \eta_n + \sum_{n=1}^N (-1)^{N-n} H_n x^{-n}.
\end{align}
Scaling $y \to (-1)^N \L^{-N}x^{\frac{N}{2}} y$ and then multiplying $(-1)^N x^{\frac{N}{2}}$ recovers \eqref{eq:spec-A}. 
The Newton polygon is represented in Fig.~\ref{fig:Newton polygon}. 

Now let us write the spectral curve in terms of the well-known $2 \times 2$ Lax operator formalism. We consider the Baker-Akhiezer function $\psi=(\psi_1,\dots,\psi_N)^T \in \BC^N$ of the Kasteleyn matrix $\fD \psi = 0$. This gives us $N$ second degree difference equations: 
\begin{align}
     (\eta_n x^{-1}-\eta^{-1}_n) \psi_n + y^{-\d_{n,1}}\L\xi_n \psi_{n-1} + y^{\d_{n,N}}\L \xi^{-1}_n x^{-1}\psi_{n+1} = 0~,~ n=1,\dots,N
\end{align}
with $\psi_{-1} = \psi_N$, $\psi_{N+1}=\psi_1$. 
The Lax matrix is obtained by rewriting the degree two difference equations into degree one matrix difference equations: 
\begin{align}
\begin{split}
    \frac{\L}{\sqrt{x}}\xi_n^{-1} \begin{pmatrix}
        y^{\d_{n,N}}\psi_{n+1} \\ \sqrt{x} \psi_n 
    \end{pmatrix} 
    & = \begin{pmatrix}
        \sqrt{x}\eta_n^{-1} - \frac{\eta_n}{\sqrt{x}} & -\L \xi_n \\ 
        \L \xi_n^{-1} & 0 
    \end{pmatrix} \begin{pmatrix}
        \psi_n \\ y^{-\d_{n,1}} \sqrt{x} \psi_{n-1}
    \end{pmatrix} \\
    & = L(x;\xi_n,\eta_n) \begin{pmatrix}
        \psi_n \\ y^{-\d_{n,1}} \sqrt{x} \psi_{n-1}
    \end{pmatrix}
\end{split}
\end{align}
We recover the Lax matrix in \eqref{def:Lax} with the identification between the Darboux coordinates and the canonical variables by
\begin{align}
    \xi_n = e^{q_n}~,~ \eta_n = e^{\frac{p_n}{2}}~.
\end{align}

\subsection{Dimer graph for reflective boundary} \label{sec:dimer-bdy}

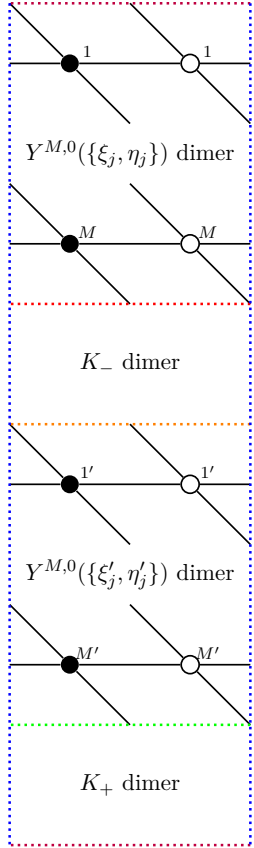
\begin{figure}[h!] \centering
\scalebox{.8}{
\begin{tikzpicture}[square/.style={regular polygon,regular polygon sides=4},decoration={markings, 
    mark= at position 0.5 with {\arrow[scale=1.3]{stealth}}}
 ]
    \node at (-4,-2) [circle,fill,inner sep=3pt] (b1) {};
    \node at (-3.7,-1.8) {${}_{1}$};
    \node at (-4,-5) [circle,fill,inner sep=3pt] (bM) {};
    \node at (-3.7,-4.8) {${}_{M}$};

    \node at (-4,-9) [circle,fill,inner sep=3pt] (b1') {};
    \node at (-3.7,-8.8) {${}_{1'}$};
    \node at (-4,-12) [circle,fill,inner sep=3pt] (bM') {};
    \node at (-3.7,-11.8) {${}_{M'}$};

    \node at (-2,-2) [style={circle,draw},thick,inner sep=3pt] (w1) {};
    \node at (-1.7,-1.8) {${}_{1}$};
    \node at (-2,-5) [style={circle,draw},thick,inner sep=3pt] (wM) {};
    \node at (-1.7,-4.8) {${}_{M}$};

    \node at (-2,-9) [style={circle,draw},thick,inner sep=3pt] (w1') {};
    \node at (-1.7,-8.8) {${}_{1'}$};
    \node at (-2,-12) [style={circle,draw},thick,inner sep=3pt] (wM') {};
    \node at (-1.7,-11.8) {${}_{M'}$};

    \draw[thick] (b1) -- (w1);
    \draw[thick] (-5,-2) -- (b1);
    \draw[thick] (-5,-1) -- (b1);
    \draw[thick] (-3,-3) -- (b1);
    \draw[thick] (w1) -- (-1,-2);
    \draw[thick] (w1) -- (-3,-1);
    \draw[thick] (w1) -- (-1,-3);

        
    \draw[thick] (bM) -- (wM);
    \draw[thick] (-5,-5) -- (bM);
    \draw[thick] (-5,-4) -- (bM);
    \draw[thick] (-3,-6) -- (bM);
    \draw[thick] (wM) -- (-1,-5);
    \draw[thick] (wM) -- (-3,-4);
    \draw[thick] (wM) -- (-1,-6);
    
    \draw[thick] (b1') -- (w1');
    \draw[thick] (-5,-9) -- (b1');
    \draw[thick] (-5,-8) -- (b1');
    \draw[thick] (-3,-10) -- (b1');
    \draw[thick] (w1') -- (-1,-9);
    \draw[thick] (w1') -- (-3,-8);
    \draw[thick] (w1') -- (-1,-10);

    
    \draw[thick] (bM') -- (wM');
    \draw[thick] (-5,-12) -- (bM');
    \draw[thick] (-5,-11) -- (bM');
    \draw[thick] (-3,-13) -- (bM');
    \draw[thick] (wM') -- (-1,-12);
    \draw[thick] (wM') -- (-3,-11);
    \draw[thick] (wM') -- (-1,-13);

    \draw[dotted,very thick,purple] (-1,-1) -- (-5,-1);
    \draw[dotted,very thick,red] (-1,-6) -- (-5,-6);
    \draw[dotted,very thick,orange] (-1,-8) -- (-5,-8);
    \draw[dotted,very thick,green] (-1,-13) -- (-5,-13);
    \draw[dotted,very thick,purple] (-1,-15) -- (-5,-15);
    \draw[dotted,very thick,blue] (-5,-1) -- (-5,-15);
    \draw[dotted,very thick,blue] (-1,-1) -- (-1,-15);


    \node at (-3,-7) {$K_-$ dimer};
    \node at (-3,-14) {$K_+$ dimer};

    \node at (-3,-3.5) {$Y^{M,0}(\{\xi_j,\eta_j\})$ dimer};
    \node at (-3,-10.5) {$Y^{M,0}(\{\xi_j',\eta_j'\})$ dimer};
    
\end{tikzpicture}
}
\caption{The general structure of RTC with reflective boundary. The horizontal purple dotted line on the top and the bottom are identified, as well as the vertical dotted lines (blue) on the left and right. In turn, the bipartite graph is drawn on a torus. }\label{fig:reflective RTC}
\end{figure}

The monodromy matrix of RTC with reflective boundary \eqref{eq:monodromy-reflection} is built by four parts: Two reflection matrices and two type A RTC in between. 
\begin{align}\label{eq:L-minus}
\begin{split}
    t_M^{-}(x) & = \s ~t^T_M(x^{-1}) \s^{-1} = (-\s L(x,\xi_1^{-1},\eta_1^{-1}) \s^{-1}) \cdots (-\s L(x,\xi_M^{-1},\eta_M^{-1}) \s^{-1}) \\
    & = (-1)^N \s L(x,\xi_1^{-1},\eta_1^{-1}) \cdots L(x,\xi_M^{-1},\eta_M^{-1}) \s^{-1}
\end{split}
\end{align}
The dimer graph for RTC with reflective boundaries is made up of four parts, the same as the monodromy matrix. The dimer graph is a bonding of two $Y^{M,0}$ models with graph for reflective boundary for $K_\pm$ at the two ends. 
\begin{align}
    M = N - \# \text{ of type D boundary}.
\end{align}
Each $Y^{M,0}$ dimer graph has $2M$ Darboux coordinates $\{\xi_n,\eta_n\}$ and $\{\xi_n',\eta'_n\}$, $n=1,\dots,M$, respectively. See Fig,~\ref{fig:reflective RTC}. 

Let $k_+$ and $k_-$ denote the number of black/white nodes on a graph for the two reflective boundaries, respectively. The Kasteleyn matrix $\fD \in \text{End}(\BC^M \otimes \BC^{k_-} \otimes \BC^M \otimes \BC^{k_+})$ of a RTC with $K_\pm$ reflection matrices is a $2M+k_++k_-$ square weighted adjacency matrix. 
Its Baker-Akhiezer function is a $(2N+k_++k_-)\times 1$ vector
\begin{align}
    \fD \Psi = 0~,~ \Psi = \sum_{j=1}^M \psi_j e_j + \sum_{j=1}^M \psi_j' e_{j}' + \sum_{i=1}^{k_-}\tilde\psi^-_i \tilde{e}^-_i + + \sum_{i=1}^{k_+}\tilde\psi^+_i \tilde{e}^+_i
\end{align}
here $e_j$, $e_j'$, $\tilde{e}_i^\pm$ are the basis vector of $\BC^M \otimes \BC^{k_-} \otimes \BC^M \otimes \BC^{k_+}$. 

The edge variables on the two $Y^{M,0}$ dimer graphs are identified by the folding procedure as follows:
\begin{align}\label{eq:folding}
    \xi_j' = \xi_{M+1-j}^{-1}~,~ \eta_j' = \eta_{M+1-j}^{-1}~,~ n=1,\dots,M~.
\end{align}

Here we give the detail structure for the reflective boundaries for the various reflection matrices $K_\pm$ mentioned in Section~\ref{sec:RTC-bdy}

\paragraph{Type C boundary}
If an RTC has a type C reflective boundary \eqref{eq:K-C}, the two $Y^{M,0}$ dimer graphs are directly connected without an additional structure. 

\paragraph{Type B-1 boundary}

\begin{figure}[h!] \centering
\scalebox{.8}{
\begin{tikzpicture}[square/.style={regular polygon,regular polygon sides=4},decoration={markings, 
    mark= at position 0.5 with {\arrow[scale=1.3]{stealth}}}
 ]
     \node at (-4,-2) [circle,fill,inner sep=3pt] (b1) {};
     \node at (-3.7,-1.8) {${}_{\tilde1}$};
     \node at (-4,-0) [circle,fill,inner sep=3pt] (b0) {};
     \node at (-3.7,.2) {${}_{M}$};
     \node at (-4,-4) [circle,fill,inner sep=3pt] (b2) {};
     \node at (-3.7,-3.8) {${}_{1'}$};

    \node at (-2,-2) [style={circle,draw},thick,inner sep=3pt] (w1) {};
    \node at (-1.7,-1.8) {${}_{\tilde1}$} ;
    \node at (-2,-0) [style={circle,draw},thick,inner sep=3pt] (w0) {};
    \node at (-1.7,.2) {${}_{M}$} ;
    \node at (-2,-4) [style={circle,draw},thick,inner sep=3pt] (w2) {};
    \node at (-1.7,-3.8) {${}_{1'}$} ;

    \draw[postaction={decorate},thick] (w1)--(b1);
    \draw[postaction={decorate},thick] (w1)--(b0);
    \draw[postaction={decorate},thick] (w1)--(-1,-2);
    \draw[postaction={decorate},thick] (w1)--(-1,-3);
    \draw[postaction={decorate},thick] (-5,-3)--(b2);

    \draw[thick] (-5,-1)--(b1);
    \draw[thick] (w2)--(b1);
    \draw[postaction={decorate},thick] (-5,-2)--(b1);

    \draw[thick] (w0)--(b0);
    \draw[thick] (w2)--(b2);
    \draw[thick] (w0)--(-1,0);
    \draw[thick] (w2)--(-1,-4);
    \draw[thick] (-5,0)--(b0);
    \draw[thick] (-5,-4)--(b2);
    \draw[thick] (w0)--(-1,-1);

    \draw[dotted,very thick,red] (-1,-1) -- (-5,-1);
    \draw[dotted,very thick,orange] (-1,-3) -- (-5,-3);
    \draw[dotted,very thick,blue] (-5,-5) -- (-5,1);
    \draw[dotted,very thick,blue] (-1,1) -- (-1,-5);


    \node[right] at (-3,-.7) {$\L\tilde\xi$};
    \node[right] at (-1,-3) {$\L\tilde\xi^{-1}$};
    \node[left] at (-5,-3) {$\L\tilde\xi^{-1}$};

    \node[right] at (-1,-2) {$\tilde\eta$};
    \node[left] at (-5,-2) {$\tilde\eta$};
    \node at (-3.1,-1.7) {$-\tilde\eta^{-1}$};


    \node at (-3,1) {$\uparrow$};
    \node at (-3,-5) {$\downarrow$};
    \node at (-3,1.5) {Connect to first $Y^{M,0}$ dimer.};
    \node at (-3,-5.5) {Connect to second $Y^{M,0}$ dimer.};
    
\end{tikzpicture}
}
\caption{The dimer graph of type B-1 reflective boundary. The vertical dotted lines (blue) on the left and right are identified.}\label{fig:B1 graph}
\end{figure}
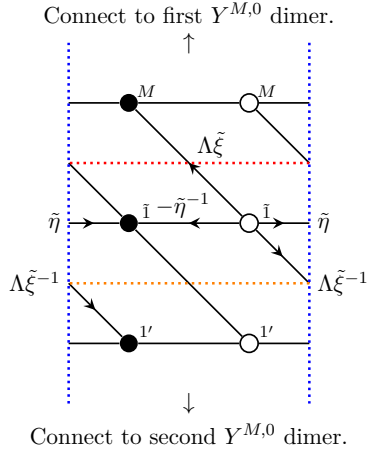

The graph for the type B-1 boundary is a $Y^{1,0}$ model. See Fig.~\ref{fig:B1 graph}. 

The Kasteleyn matrix for the $Y^{1,0}$ model in Fig.~\ref{fig:B1 graph} is
\begin{align}
    \left( 
    \begin{array}{c | c c c }
         & b_M & \tilde{b} & b_1' \\
         \hline
         \tilde{w} & \L^{\frac12}\tilde\xi & x^{-1}\tilde\eta-\tilde\eta^{-1} & \L^{\frac12}\tilde\xi^{-1} x^{-1}
    \end{array}
    \right)
\end{align}
Here $b_M$ is the last black node of the $Y^{M,0}$ graph on the top. $b_1'$ is the first black node of the $Y^{M,0}$ on the bottom. 

The Baker-Akhiezer function of the Kasteleyn matrix gives
\begin{subequations}
\begin{align}
    & \L^{\frac12}\tilde\xi \psi_N + (x^{-1}\tilde\eta-\tilde\eta^{-1}) \tilde\psi + \L^{\frac12}\tilde\xi^{-1} x^{-1}\psi_1' = 0
\end{align}
\end{subequations}
The degree-two difference equation above is organized into degree-one difference matrix equation. 
\begin{align}
    \frac{\L}{x^{\frac12}} \begin{pmatrix}
        \psi_1' \\ x^{\frac12}\tilde\psi
    \end{pmatrix} = L(x;\tilde\xi,\tilde\eta) \begin{pmatrix}
        \tilde\psi \\ x^{\frac12}\psi_N
    \end{pmatrix}
\end{align}
The type B-1 reflective boundary \eqref{eq:K-B1} comes from freezing Darboux coordinates.
\begin{align}
    K^B_+(x) = \s \frac{L(x,\tilde\xi=1,\tilde\eta=1)}{x^{\frac12}-x^{-\frac12}}~,~ K_-^B(x) = \frac{L(x;\tilde\xi=1,\tilde\eta=1)}{x^{\frac12}-x^{-\frac12}}\s^{-1}~.
\end{align}
The dimer graph in Fig.~\ref{fig:B1 graph} introduces one more particle into the system whose position and conjugate momentum are $(\tilde\xi,\tilde\eta)$. Physically, freezing the Darboux coordinate fixes this particles to be immobile and serves as a boundary of the system. 

Combining the two $Y^{M,0}$ dimer graph across a type B-1 boundary creates a $Y^{2M+1,0}$ model graph, with the canonical/Daroux coordinates in the very center $\xi_{M+1}|_{Y^{2M+1,0}}=1=\eta_{M+1}|_{Y^{2M+1,0}}$ level frozen. 

\begin{figure}[h!] \centering
\scalebox{.8}{
\begin{tikzpicture}[square/.style={regular polygon,regular polygon sides=4},decoration={markings, 
    mark= at position 0.5 with {\arrow[scale=1.3]{stealth}}}
 ]
     \node at (-4,-2) [circle,fill,inner sep=3pt] (b1) {};
     \node at (-3.7,-1.8) {${}_{\tilde1}$};
      \node at (-4,-4) [circle,fill,inner sep=3pt] (b2) {};
     \node at (-3.7,-3.8) {${}_{\tilde2}$} ;
    \node at (-4,-0) [circle,fill,inner sep=3pt] (b0) {};
     \node at (-3.7,.2) {${}_{M}$};
     \node at (-4,-6) [circle,fill,inner sep=3pt] (b3) {};
     \node at (-3.7,-5.8) {${}_{1'}$};

    \node at (-2,-2) [style={circle,draw},thick,inner sep=3pt] (w1) {};
    \node at (-1.7,-1.8) {${}_{\tilde1}$} ;
    \node at (-2,-4) [style={circle,draw},thick,inner sep=3pt] (w2) {};
    \node at (-1.7,-3.8) {${}_{\tilde2}$} ;
    \node at (-2,-0) [style={circle,draw},thick,inner sep=3pt] (w0) {};
    \node at (-1.7,.2) {${}_{M}$} ;
    \node at (-2,-6) [style={circle,draw},thick,inner sep=3pt] (w3) {};
    \node at (-1.7,-5.8) {${}_{1'}$} ;

    \draw[postaction={decorate},thick] (w1)--(b1);
    \draw[postaction={decorate},thick] (w2)--(b2);
    \draw[postaction={decorate},thick] (w2)--(b1);

    \draw[postaction={decorate},thick] (w1)--(b0);
    \draw[postaction={decorate},thick] (w1)--(-1,-2);
    \draw[postaction={decorate},thick] (w1)--(-1,-3);
    \draw[postaction={decorate},thick] (w2)--(-1,-4);
    \draw[postaction={decorate},thick] (w2)--(-1,-5);

    \draw[thick] (-5,-1)--(b1);
    \draw[postaction={decorate},thick] (-5,-2)--(b1);
    \draw[postaction={decorate},thick] (-5,-3)--(b2);
    \draw[postaction={decorate},thick] (-5,-4)--(b2);
    \draw[thick] (w3)--(b2);

    \draw[thick] (w0) -- (b0);
    \draw[thick] (w0) -- (-1,-1);
    \draw[thick] (w0) -- (-1,0);
    \draw[thick] (-5,0) -- (b0);

    \draw[thick] (w3) -- (b3);
    \draw[thick] (w3) -- (-1,-6);
    \draw[postaction={decorate},thick] (-5,-5) -- (b3);
    \draw[thick] (-5,-6) -- (b3);

    \draw[dotted,very thick,red] (-1,-1) -- (-5,-1);
        \draw[dotted,very thick,orange] (-1,-5) -- (-5,-5);
       \draw[dotted,very thick,blue] (-5,0.5) -- (-5,-6.5);
        \draw[dotted,very thick,blue] (-1,0.5) -- (-1,-6.5);


    \node[right] at (-3,-.7) {$\L^{\frac12}\tilde\xi_1$};
    \node[right] at (-1,-3) {$\L^{\frac12}\tilde\xi_1^{-1}$};
    \node[left] at (-5,-3) {$\L^{\frac12}\tilde\xi_1^{-1}$};
    \node[right] at (-3,-2.8) {$\L^{\frac12}\tilde\xi_2$};
    \node[right] at (-1,-5) {$\L^{\frac12}\tilde\xi_2^{-1}$};

    \node[right] at (-1,-2) {$\tilde\eta_1$};
    \node[left] at (-5,-2) {$\tilde\eta_1$};
    \node[right] at (-1,-4) {$\tilde\eta_2$};
    \node[left] at (-5,-4) {$\tilde\eta_2$};
    \node at (-3.1,-1.7) {$-\tilde\eta_1^{-1}$};
    \node at (-3,-4.3) {$-\tilde\eta_2^{-1}$};


    \node at (-3,.5) {$\uparrow$};
    \node at (-3,-6.5) {$\downarrow$};
    \node at (-3,1) {Connect to first $Y^{M,0}$ dimer.};
    \node at (-3,-7) {Connect to second $Y^{M,0}$ dimer.};
    
\end{tikzpicture}
}
\caption{The dimer graph of type B-2 reflective boundary. The vertical dotted lines (blue) on the left and right are identified. }\label{fig:B2 graph}
\end{figure}
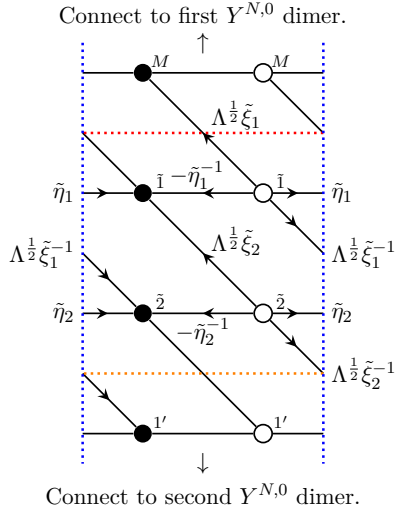

\paragraph{Type B-2 boundary}
The reflection matrices for type B2 boundary \eqref{eq:K-B2} are
\begin{align}
    \bar{K}^{B}_+(x,\k_+) = \frac{1}{x-x^{-1}} \begin{pmatrix}
        \L x^{-\frac{\k_+}{2}} & 0 \\ x^{-1}-x & \L x^{-\frac{\k_+}{2}}
    \end{pmatrix}~,~ \bar{K}^{B}_- (x,\k_-) = \frac{1}{x-x^{-1}} \begin{pmatrix}
        -\L x^{-\frac{\k_-}{2}} & x^{-1}-x \\ 0 & -\L x^{\frac{\k_-}{2}}
    \end{pmatrix}~.
\end{align}
The graph for a type B-2 boundary is the $Y^{2,0}$ model with a different weight on the edge. See Fig.~\ref{fig:B2 graph}. 

The Kasteleyn matrix for the $Y^{2,0}$ model in Fig.~\ref{fig:B2 graph} is
\begin{align}
    \left( 
    \begin{array}{c | c c c c}
         & b_M & \tilde{b}_1 & \tilde{b}_2 & b_1' \\
         \hline
         \tilde{w}_1 & \L^{\frac12}\tilde\xi_1 & x^{-1}\tilde\eta_1-\tilde\eta_1^{-1} & \L^{\frac12}\tilde\xi_1^{-1} x^{-1} & 0 \\
         \tilde{w}_2 & 0 & \L^{\frac12} \xi_2 & x^{-1}\tilde\eta_2-\tilde\eta_2^{-1} & \L^{\frac12}\tilde\xi_2^{-1} x^{-1}  
    \end{array}
    \right)
\end{align}
Here $b_M$ is the last black node of the $Y^{M,0}$ graph on the top. $b_1''$ is the first black node of the $Y^{M,0}$ on the bottom. 

The Baker-Akhiezer function of the Kasteleyn matrix gives
\begin{subequations}
\begin{align}
    & \L^{\frac12}\tilde\xi_1 \psi_M + (x^{-1}\tilde\eta_1-\tilde\eta_1^{-1}) \tilde\psi_1 + \L^{\frac12}\tilde\xi_1^{-1} x^{-1} \tilde\psi_2 = 0
    & \L^{\frac12}\tilde\xi_2 \tilde\psi_1 + (x^{-1}\tilde\eta_2-\tilde\eta_2^{-1}) \psi_2 + \L^{\frac12}\tilde\xi_1^{-1} x^{-1} \psi_1' = 0
\end{align}
\end{subequations}
The degree-two difference equations above are organized into degree-one difference matrix equation
\begin{align}
\begin{split}
    \frac{\L}{x^{\frac12}} \begin{pmatrix}
        \psi_1' \\ x^{\frac12}\tilde\psi_2
    \end{pmatrix} 
    & = \begin{pmatrix}
        \L^{\frac12} \frac{\sqrt{x}}{\tilde\xi_{2}\tilde\eta_{1}}  - \L^{\frac12} \frac{\tilde\eta_{1}}{\tilde\xi_{2}\sqrt{x}} & -\L \frac{\tilde\xi_{1}}{\tilde\xi_{2}} \\
        -\frac{x}{\tilde\eta_{2}\tilde\eta_{1}} + \frac{\tilde\eta_{1}}{\eta_{2}} + \frac{\tilde\eta_{2}}{\tilde\eta_{1}} - \frac{\tilde\eta_{1}\tilde\eta_{2}}{x} + \L \frac{\tilde\xi_{2}}{\tilde\xi_{1}} & \L^{\frac12} \sqrt{x}\frac{\tilde\xi_{1}}{\tilde\eta_{2}} - \L^{\frac12} \frac{\tilde\xi_{1}\tilde\eta_{2}}{\sqrt{x}}
    \end{pmatrix} 
    \begin{pmatrix}
        \tilde\psi_1 \\ x^{\frac12} \psi_M
    \end{pmatrix} \\
    & := \bar{K}^B (x;\tilde\xi_1,\tilde\xi_2,\tilde\eta_1,\tilde\eta_2) \begin{pmatrix}
        \tilde\psi_1 \\ x^{\frac12} \psi_M
    \end{pmatrix}
\end{split}
\end{align}
To recover the reflection matrix, we freeze the Darboux coordinates by
\begin{align}
    \tilde\xi_1 = e^{-\k \frac{\pi i}{4}} \L^{\frac12} r^{-1}~,~ \tilde\xi_2 = e^{-\k \frac{\pi i}{4}} \L^{-\frac12}r~,~ \tilde\eta_1 = e^{-\frac{3\pi i}{4}} r^{\k}~,~ \tilde\eta_2 = e^{\frac{\pi i}{4}} r^{-\k}
\end{align}
so that 
\begin{align}
\begin{split}
    & \lim_{r\to \infty} \bar{K}^B (x;e^{-\k \frac{\pi i}{4}} \L^{\frac12} r^{-1},\xi_2=e^{-\k \frac{\pi i}{4}} \L^{-\frac12},\eta_1=e^{-\frac{3\pi i}{4}} r^{\k},\eta_2=e^{\frac{\pi i}{4}} r^{-\k}) \\
    & = i \begin{pmatrix}
        x - x^{-1} & - \L x^{\frac{\k}{2}} \\ 
        \L x^{-\frac{\k}{2}} & 0
    \end{pmatrix} = -i\s (x-x^{-\frac12}) \bar{K}^B_+(x,\k) = i (x-x^{-\frac12})\bar{K}^B_-(x,-\k) \s~.
\end{split}
\end{align}
Similar to the type B1 case, the dimer graph in Fig.~\ref{fig:B2 graph} introduces two additional particles to the physical system with positions and conjugate momentums as $(\tilde\xi_{1,2},\tilde\eta_{1,2})$. Freezing the Darboux coordinates makes these newly introduced particles immobile, serving as a boundary of the system. 

\paragraph{Type D boundary}
The dimer graph for type D boundary is explicitly constructed in \cite{Lee:2024bqg}. 
Here we give a brief review on the construction.

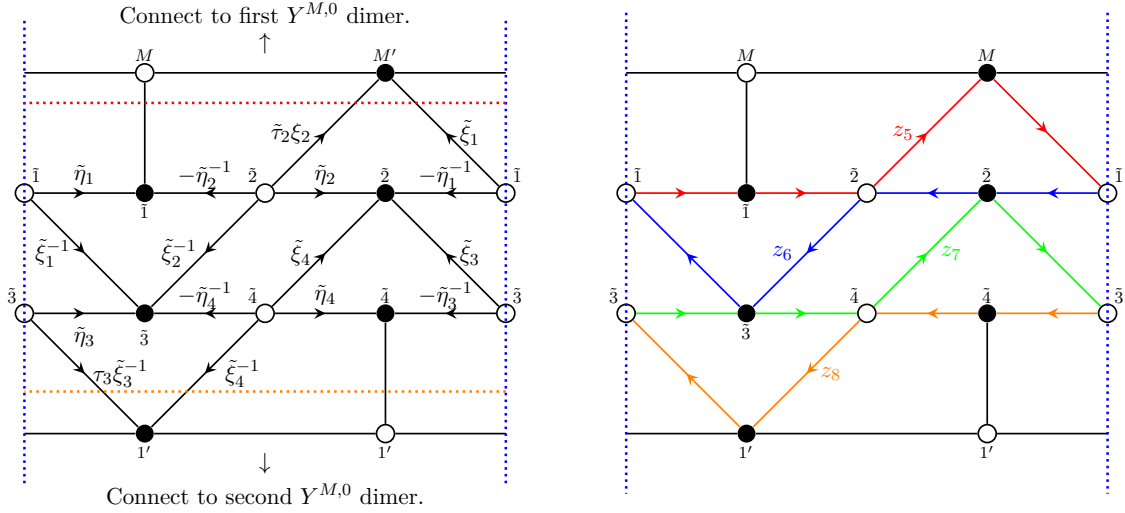
\begin{figure}[h!] \centering
\scalebox{.8}{
\begin{tikzpicture}[square/.style={regular polygon,regular polygon sides=4},decoration={markings, 
    mark= at position 0.5 with {\arrow[scale=1.3]{stealth}}}
 ]
     \node at (-2,-2) [circle,fill,inner sep=3pt] (b1) {};
     \node at (-2,-1.7) {${}_{M'}$};
     \node at (-2,-4) [circle,fill,inner sep=3pt] (b3) {};
     \node at (-2,-3.7) {${}_{\tilde2}$};
     \node at (-6,-4) [circle,fill,inner sep=3pt] (b2) {};
     \node at (-6,-4.3) {${}_{\tilde1}$};
     \node at (-2,-6) [circle,fill,inner sep=3pt] (b5) {};
     \node at (-2,-5.7) {${}_{\tilde4}$};
     \node at (-6,-6) [circle,fill,inner sep=3pt] (b4) {};
     \node at (-6,-6.4) {${}_{\tilde3}$};
     \node at (-6,-8) [circle,fill,inner sep=3pt] (b6) {};
     \node at (-6,-8.3) {${}_{1'}$};

    \node at (-6,-2) [style={circle,draw},thick,inner sep=3pt] (w1) {};
    \node at (-6,-1.7) {${}_{M}$} ;
    \node at (-4,-4) [style={circle,draw},thick,inner sep=3pt] (w2) {};
    \node at (-4.2,-3.7) {${}_{\tilde2}$} ;
    \node at (-0,-4) [style={circle,draw},thick,inner sep=3pt] (w31) {};
    \node at (.2,-3.7) {${}_{\tilde1}$} ;
    \node at (-8,-4) [style={circle,draw},thick,inner sep=3pt] (w32) {};
    \node at (-7.8,-3.7) {${}_{\tilde1}$} ;
    \node at (-4,-6) [style={circle,draw},thick,inner sep=3pt] (w5) {};
    \node at (-4.2,-5.7) {${}_{\tilde4}$} ;
    \node at (-0,-6) [style={circle,draw},thick,inner sep=3pt] (w41) {};
    \node at (-8.2,-5.7) {${}_{\tilde3}$} ;
    \node at (-8,-6) [style={circle,draw},thick,inner sep=3pt] (w42) {};
    \node at (.2,-5.7) {${}_{\tilde3}$} ;
    \node at (-2,-8) [style={circle,draw},thick,inner sep=3pt] (w6) {};
    \node at (-2,-8.3) {${}_{1'}$} ;


    \draw[thick] (w1)--(b1);
    \draw[thick] (-0,-2)--(b1);
    \draw[thick] (w1)--(-8,-2);
    \draw[thick] (w1)--(b2);

    \draw[postaction={decorate},thick] (w2)--(b1);
    \draw[postaction={decorate},thick] (w2)--(b2);
    \draw[postaction={decorate},thick] (w2)--(b3);
    \draw[postaction={decorate},thick] (w2)--(b4);

    \draw[postaction={decorate},thick] (w31)--(b1);
    \draw[postaction={decorate},thick] (w31)--(b3);
    \draw[postaction={decorate},thick] (w32)--(b2);
    \draw[postaction={decorate},thick] (w32)--(b4);
    
    \draw[postaction={decorate},thick] (w41)--(b3);
    \draw[postaction={decorate},thick] (w41)--(b5);
    \draw[postaction={decorate},thick] (w42)--(b4);
    \draw[postaction={decorate},thick] (w42)--(b6);

    \draw[postaction={decorate},thick] (w5)--(b3);
    \draw[postaction={decorate},thick] (w5)--(b4);
    \draw[postaction={decorate},thick] (w5)--(b5);
    \draw[postaction={decorate},thick] (w5)--(b6);

    \draw[thick] (w6)--(b5);
    \draw[thick] (w6)--(b6);
    \draw[thick] (w6)--(0,-8);
    \draw[thick] (-8,-8)--(b6);
    
    \node (t1) at (0,-1) {};
    \node (t2) at (-8,-1) {};

    \node (bb1) at (0,-9) {};
    \node (bb2) at (-8,-9) {};
    
    \draw[dotted,very thick,blue] (t1) -- (bb1);
    \draw[dotted,very thick,blue] (t2) -- (bb2);

    \draw[dotted,very thick,red] (0,-2.5) -- (-8,-2.5);
    \draw[dotted,very thick,orange] (0,-7.3) -- (-8,-7.3);


    
    \node at (-3,-3.7) {$\tilde\eta_2$};
    \node at (-5,-3.7) {$-\tilde\eta_2^{-1}$};
    \node[left] at (-5,-5) {$\tilde\xi_2^{-1}$};
    \node[left] at (-3.1,-3) {$\tilde\t_2\xi_2$};

    \node at (-1,-3.7) {$-\tilde\eta_1^{-1}$};
    \node at (-7,-3.7) {$\tilde\eta_1$};
    \node[right] at (-.9,-3) {$\tilde\xi_1$};
    \node[left] at (-7.1,-5) {$\tilde\xi_1^{-1}$};

    \node at (-1,-5.7) {$-\tilde\eta_3^{-1}$};
    \node at (-7,-6.4) {$\tilde\eta_3$};
    \node[right] at (-.9,-5) {$\tilde\xi_3$};
    \node[right] at (-7,-7) {$\t_3\tilde\xi_3^{-1}$};

    \node at (-3,-5.7) {$\tilde\eta_4$};
    \node at (-5,-5.7) {$-\tilde\eta_4^{-1}$};
    \node[right] at (-4.8,-7) {$\tilde\xi_4^{-1}$};
    \node[left] at (-3.1,-5) {$\tilde\xi_4$};



     \node at (8,-2) [circle,fill,inner sep=3pt] (b1z) {};
     \node at (8,-1.7) {${}_{M}$};
     \node at (8,-4) [circle,fill,inner sep=3pt] (b3z) {};
     \node at (8,-3.7) {${}_{\tilde2}$};
     \node at (4,-4) [circle,fill,inner sep=3pt] (b2z) {};
     \node at (4,-4.3) {${}_{\tilde1}$};
     \node at (8,-6) [circle,fill,inner sep=3pt] (b5z) {};
     \node at (8,-5.7) {${}_{\tilde4}$};
     \node at (4,-6) [circle,fill,inner sep=3pt] (b4z) {};
     \node at (4,-6.3) {${}_{\tilde3}$};
     \node at (4,-8) [circle,fill,inner sep=3pt] (b6z) {};
     \node at (4,-8.3) {${}_{1'}$};
     
    \node at (4,-2) [style={circle,draw},thick,inner sep=3pt] (w1z) {};
    \node at (4,-1.7) {${}_{M}$} ;
    \node at (6,-4) [style={circle,draw},thick,inner sep=3pt] (w3z) {};
    \node at (5.8,-3.7) {${}_{\tilde2}$} ;
    \node at (10,-4) [style={circle,draw},thick,inner sep=3pt] (w21z) {};
    \node at (10.2,-3.7) {${}_{\tilde1}$} ;
    \node at (2,-4) [style={circle,draw},thick,inner sep=3pt] (w22z) {};
    \node at (2.2,-3.7) {${}_{\tilde1}$} ;
    \node at (6,-6) [style={circle,draw},thick,inner sep=3pt] (w5z) {};
    \node at (5.8,-5.7) {${}_{\tilde4}$} ;
    \node at (10,-6) [style={circle,draw},thick,inner sep=3pt] (w41z) {};
    \node at (1.8,-5.7) {${}_{\tilde3}$} ;
    \node at (2,-6) [style={circle,draw},thick,inner sep=3pt] (w42z) {};
    \node at (10.2,-5.7) {${}_{\tilde3}$} ;
    \node at (8,-8) [style={circle,draw},thick,inner sep=3pt] (w6z) {};
    \node at (8,-8.3) {${}_{1'}$} ;

    \draw[postaction={decorate},thick,red] (w22z)--(b2z);
    \draw[postaction={decorate},thick,red] (b2z)--(w3z);
    \draw[postaction={decorate},thick,red] (w3z)--(b1z);
    \draw[postaction={decorate},thick,red] (b1z)--(w21z);

    \draw[postaction={decorate},thick,blue] (w21z)--(b3z);
    \draw[postaction={decorate},thick,blue] (b3z)--(w3z);
    \draw[postaction={decorate},thick,blue] (w3z)--(b4z);
    \draw[postaction={decorate},thick,blue] (b4z)--(w22z);

    \draw[postaction={decorate},thick,green] (w42z)--(b4z);
    \draw[postaction={decorate},thick,green] (b4z)--(w5z);
    \draw[postaction={decorate},thick,green] (w5z)--(b3z);
    \draw[postaction={decorate},thick,green] (b3z)--(w41z);

    \draw[postaction={decorate},thick,orange] (w41z)--(b5z);
    \draw[postaction={decorate},thick,orange] (b5z)--(w5z);
    \draw[postaction={decorate},thick,orange] (w5z)--(b6z);
    \draw[postaction={decorate},thick,orange] (b6z)--(w42z);

    \draw[thick] (b1z) -- (w1z);
    \draw[thick] (b2z) -- (w1z);
    \draw[thick] (b6z) -- (w6z);
    \draw[thick] (b5z) -- (w6z);

    \draw[thick] (b1z) -- (10,-2);
    \draw[thick] (2,-2) -- (w1z);
    \draw[thick] (b6z) -- (2,-8);
    \draw[thick] (10,-8) -- (w6z);

    \node at (6.6,-3) {\color{red}$z_5$};
    \node at (4.6,-5) {\color{blue}$z_6$};
    \node at (7.4,-5) {\color{green}$z_7$};
    \node at (5.4,-7) {\color{orange}$z_8$};

    \draw[dotted,very thick,blue] (2,-1) -- (2,-9);
    \draw[dotted,very thick,blue] (10,-1) -- (10,-9);

    
    \node at (-4,-1.5) {$\uparrow$};
    \node at (-4,-8.5) {$\downarrow$};
    \node at (-4,-1) {Connect to first $Y^{M,0}$ dimer.};
    \node at (-4,-9) {Connect to second $Y^{M,0}$ dimer.};
    
\end{tikzpicture}
}
\caption{The dimer for type D reflective boundary. The vertical dotted lines (blue) on the left and right are identified. \\
Left: gauged edge variables. 
Right: zigzag loops generated by the double impurity. }\label{fig:double impurity}
\end{figure}

At the position where the reflection matrix $K^D(x;q,p)$ is located, we modify the dimer graph as shown in Fig.~\ref{fig:double impurity}. This modification is known as \emph{double impurity} in \cite{Lee:2024bqg}.   

The submatrix for the double impurity in Fig.~\ref{fig:double impurity} is
\begin{align}
\begin{split}
    \left( 
    \begin{array}{c|c c c c c c}
         & b_N & \tilde{b}_1 & \tilde{b}_2 & \tilde{b}_3 & \tilde{b}_4 & b_1' \\
         \hline
        w_1 & \tilde\xi_1 x^{\frac12} & \tilde\eta_1 x^{-\frac12} & -\tilde\eta_1^{-1}x^{\frac12} & \tilde\xi_1^{-1} x^{-\frac12} & & \\
        w_2 & \t_2\tilde\xi_2 & -\tilde\eta_2^{-1} & \tilde\eta_2 & \xi_2^{-1} & & \\
        w_3 & & & \tilde\xi_3 x^{\frac12} & \tilde\eta_3 x^{-\frac12} & -\tilde\eta_3^{-1}x^{\frac12} & \t_3\tilde\xi_3^{-1} x^{-\frac12} \\
        w_4 & & & \xi_4 & -\tilde\eta_4^{-1} & \tilde\eta_4 & \xi_4^{-1} \\
    \end{array}
    \right)~.
\end{split}
\end{align}
Four zigzag loops exist in Fig.~\ref{fig:double impurity}:
\begin{align}
     z_5=\t_2\tilde\eta_1\tilde\xi_1^{-1}\tilde\eta_2\tilde\xi_2~,~z_6=\tilde\eta_1^{-1}\tilde\xi_1\tilde\eta_2^{-1}\tilde\xi_2^{-1} ~,~z_7=\tilde\eta_3\tilde\xi_3^{-1}\tilde\eta_4\tilde\xi_4 ~,~ z_8=\tilde\t_3^{-1}\tilde\eta_3^{-1}\tilde\xi_3\tilde\eta_4^{-1}\xi_4^{-1} ~ . 
\end{align}
The Baker-Akhiezer function of the Kasteleyn matrix around the impurity gives
\begin{subequations}\label{eq:D-BA}
\begin{align}
    & \tilde\xi_1 x^{\frac12} \psi_M + \tilde\eta_1 x^{-\frac12} \tilde\psi_1 - \tilde\eta_1^{-1} x^{\frac12} \tilde\psi_2 + \tilde\xi_1^{-1} x^{-\frac12} \tilde\psi_3 = 0 ~,\\
    & \t_2\tilde\xi_2 \psi_M - \tilde\eta_2 \tilde\psi_1 + \tilde\eta_2 \tilde\psi_2 + \tilde\xi_2^{-1} \tilde\psi_3 = 0 ~,\\
    & \tilde\xi_3 x^{\frac12} \tilde\psi_2 + \tilde\eta_3 x^{-\frac12} \tilde\psi_3 - \tilde\eta_3^{-1} x^{\frac12} \tilde\psi_4 + \t_3\tilde\xi_4^{-1} x^{-\frac12} \psi_1' = 0 ~,\\
    & \tilde\xi_4 \tilde\psi_2 - \tilde\eta_4^{-1} \tilde\psi_3 + \tilde\eta_4 \tilde\psi_4 + \tilde\xi_4 \psi_1' = 0 ~.
\end{align}
\end{subequations}
We set $z_6=z_7=1$, $\t_3=\t_4=-1$ and perform the folding by
\begin{align}
    \tilde\eta_1\tilde\eta_3\tilde\xi_1\tilde\xi_3 = 1~,~\tilde\eta_2\tilde\eta_4\tilde\xi_2^{-1}\tilde\xi_4^{-1} = 1~,~ \tilde\eta_1\tilde\eta_3 \tilde\xi_1^{-1}\tilde\xi_3^{-1}=-1~.
\end{align}
The relation between the Darboux coordinates and canonical coordinates is given by 
\begin{align}\label{eq:D-Darboux-canonical}
    \tilde\eta_1\tilde\xi_1^{-1} = - \frac{\sinh(p/2)}{\sinh(q)}~,~ \tilde\eta_1\tilde\eta_2\xi_1\xi_2^{-1} = \frac{\sinh(p/2+q)}{\sinh(p/2-q)}~.
\end{align}
One can check that the two functions are log-canonical. 
The reflection matrices $K^D$ are recovered from \eqref{eq:D-BA}:
\begin{align}
\begin{split}
    & (x-x^{-1}) \begin{pmatrix}
        \psi_1' \\ x^{\frac12} \tilde\psi_4
    \end{pmatrix} 
    =\s^{-1} K_+^D (x;q,p) \begin{pmatrix}
        \tilde\psi_1 \\ x^{\frac12}\psi_M
    \end{pmatrix} 
    = K_-^D (x;q,p) \s \begin{pmatrix}
        \tilde\psi_1 \\ x^{\frac12}\psi_M
    \end{pmatrix}
\end{split}
\end{align}
with the type D reflection matrix defined in \eqref{eq:K-D}. 

We want to mention that the construction of the type D boundary here is an improvement over \cite{Lee:2024bqg} since we no longer need to modify two of the face variables in Fig.~\ref{fig:double impurity} as in \cite{Lee:2024bqg}. 

\section{Examples}\label{sec:example}

\subsection{${C}_N^{(1)}$}

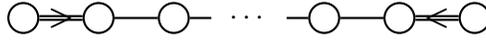
\begin{figure}[h!] \centering
\scalebox{1}{
\begin{tikzpicture}[square/.style={regular polygon,regular polygon sides=4},decoration={markings, 
    mark= at position 0.5 with {\arrow[scale=1.3]{stealth}}}
 ]
    
    \node at (-3,-2) [style={circle,draw},thick,inner sep=4pt] (w1) {};
    \node at (-2,-2) [style={circle,draw},thick,inner sep=4pt] (w2) {};
    \node at (-1,-2) [style={circle,draw},thick,inner sep=4pt] (w3) {};
    \node at (1,-2) [style={circle,draw},thick,inner sep=4pt] (w4) {};
    \node at (2,-2) [style={circle,draw},thick,inner sep=4pt] (w5) {};
    \node at (3,-2) [style={circle,draw},thick,inner sep=4pt] (w6) {};

    \draw[thick,double] (w1) -- (w2);
    \draw[thick] (w2) -- (w3);
    \draw[thick] (w3) -- (-.5,-2);
    \draw[thick] (w4) -- (.5,-2);
    \draw[thick] (w4) -- (w5);
    \draw[thick,double] (w5) -- (w6);
    \node at (0,-2) {$\cdots$};

    \node at (2.5,-2) {$\boldsymbol{<}$};
    \node at (-2.5,-2) {$\boldsymbol{>}$};
    
\end{tikzpicture}
}
\caption{Dynkin diagram for ${C}_N^{(1)}$}\label{fig:C1N-Dynkin}
\end{figure}

The Dynkin diagram of affine ${C}_N^{(1)}$ is illustrated in Fig.~\ref{fig:C1N-Dynkin}. It has type C boundary on both ends. 

The Hamilton is
\begin{align}
    H_1 = H_A + J^C_++J^C_- = H_A + \L^2 e^{2q_1} + \L^2 e^{-2q_N}~.
\end{align}
The reflection matrices are chosen based on the structure of the the Dynkin diagram of $C_N^{(1)}$ in \eqref{eq:K-C}. 
The monodromy matrix is 
\begin{align}
    T(x)|_{C^{(1)}_N} = K_+^C ~t_N(x) K^C_- t_N^-(x)~.
\end{align}
The spectral curve is given by
\begin{align}
    y + \frac{\L^{4N}}{y} = x^N + x^{-N} + \sum_{n=1}^{N} H_n (x^{N-n}+x^{n-N})~.  
\end{align}
The ${C}_N^{(1)}$ RTC shares the same toric diagram as ${A}_{2N}^{(1)}$. 
The dimer graph for $C^{(1)}_N$ RTC is simply the gluing of two $Y^{N,0}$ dimer graph, i.e. the same as $Y^{2N,0}$ dimer graph. The Darboux coordinates of the two $Y^{N,0}$ model are folded \eqref{eq:folding} and related to the canonical coordinates by
\begin{align}
    \xi_{N+1-j}^{-1} = \xi'_{j} = e^{q_j}~,~ \eta_{N+1-j}^{-1} = \eta_{j}' = e^{\frac{p_j}{2}}~,~ j=1,\dots,N.
\end{align}

\subsection{$({C}_N^{(1)})^\vee=D_{N+1}^{(2)}$}\label{sec:C-dual}

\begin{figure}[h!] \centering
\scalebox{1}{
\begin{tikzpicture}[square/.style={regular polygon,regular polygon sides=4},decoration={markings, 
    mark= at position 0.5 with {\arrow[scale=1.3]{stealth}}}
 ]
    
    \node at (-3,-2) [style={circle,draw},thick,inner sep=4pt] (w1) {};
    \node at (-2,-2) [style={circle,draw},thick,inner sep=4pt] (w2) {};
    \node at (-1,-2) [style={circle,draw},thick,inner sep=4pt] (w3) {};
    \node at (1,-2) [style={circle,draw},thick,inner sep=4pt] (w4) {};
    \node at (2,-2) [style={circle,draw},thick,inner sep=4pt] (w5) {};
    \node at (3,-2) [style={circle,draw},thick,inner sep=4pt] (w6) {};

    \draw[thick,double] (w1) -- (w2);
    \draw[thick] (w2) -- (w3);
    \draw[thick] (w3) -- (-.5,-2);
    \draw[thick] (w4) -- (.5,-2);
    \draw[thick] (w4) -- (w5);
    \draw[thick,double] (w5) -- (w6);
    \node at (0,-2) {$\cdots$};

    \node at (2.5,-2) {$\boldsymbol{>}$};
    \node at (-2.5,-2) {$\boldsymbol{<}$};
    
\end{tikzpicture}
}
\caption{Dynkin diagram for $({C}_N^{(1)})^\vee=D_{N+1}^{(2)}$}\label{fig:D2N-Dynkin}
\end{figure}
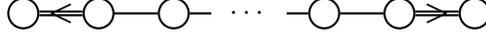

Lie algebra that are not simply laced have dual algebra with long and short roots exchanged. $(C_N^{(1)})^\vee$, aka  twisted affine Lie algebra $D^{(2)}_{N+1}$, is the Lie algebra for $Sp(N)^\vee$. 
The Dynkin diagram of $({C}_N^{(1)})^\vee$ is illustrated in Fig.~\ref{fig:D2N-Dynkin}. The Dynkin diagram has type B boundaries at both ends. 

\paragraph{Type B-1:}
The Hamiltonian is given by
\begin{align}
    H_1 = H_A + J^B_+ + J^B_- =  H_A + 2\L^2 e^{q_1} \cosh \frac{p_1}{2} + 2\L^2 e^{-q_{N}}\cosh\frac{p_N}{2}. 
\end{align}
Based on the type B boundary of the Dynkin diagram. The reflection matrices \eqref{eq:K-B1} are
\begin{align}
    K_+^B = \begin{pmatrix}
        \frac{\L}{x^{\frac12}-x^{-\frac12}} & 0 \\ -1 & \frac{\L}{x^{\frac12}-x^{-\frac12}}
    \end{pmatrix}~,~ K_-^B = \begin{pmatrix}
        - \frac{\L}{x^{\frac12}-x^{-\frac12}} & -1 \\ 0 & -\frac{\L}{x^{\frac12}-x^{-\frac12}}
    \end{pmatrix}
\end{align}

In particular the Hamiltonian equals to type A open RTC with $N+2$ particles, and having the first and $(N+2)$-th particles frozen, i.e. setting $p_0=p_{N+1}=0=q_0=q_{N+1}$. 
The monodromy matrix is
\begin{align}
    T(x)|_{(C^{(1)}_N)^\vee} = K^B_+ t_N(x) K_-^B t_N^-(x) 
\end{align}
with spectral curve
\begin{align}
    y + \frac{\L^{4N+4}}{y} \frac{x^2}{(x-1)^4} = x^{N} + x^{-N} + \sum_{n=1}^N H_n (x^{N-n}+x^{n-N}) - \frac{2\L^{2N+2}x}{(x-1)^2}. 
\end{align}
Use a combination of the $SL(2,\BZ)$ action \eqref{eq:sl2z-trans} and birational transformation \eqref{eq:bira-trans} to scale $y \to yx(x-1)^{-2}$ and multiply the spectral curve by $(x-1)^2/x$, we obtain
\begin{align}
    y + 2 \L^{2N+2} + \frac{\L^{4N+4}}{y} = \left( x-2+\frac1x \right) \left[ x^{N} + x^{-N} + \sum_{n=1}^N (-1)^n H_n (x^{N-n}+x^{n-N}) \right]
\end{align}
The dimer graph Fig.~\ref{fig:reflective RTC} is a gluing between two $Y^{N,0}$ dimer graph with two type B-1 boundary. It shares the same shape as the $Y^{2N+2,0}$ dimer. The Darboux coordinates of the two $Y^{N,0}$ model are folded \eqref{eq:folding} and related to the canonical coordinates by
\begin{align}
    \xi_{N+1-j}^{-1} = \xi'_{j} = e^{q_j}~,~ \eta_{N+1-j}^{-1} = \eta_{j}' = e^{\frac{p_j}{2}}~,~ j=1,\dots,N.
\end{align}

\paragraph{Type B-2:}\label{sec:B-2} 
The Hamiltonian is
\begin{align}
    H_1 = H_A + \bar{J}_+^B + \bar{J}^B_- = H_A + \L^2 e^{q_1+\k_-\frac{p_1}{2}} + \L^2 e^{-q_N+\k_+\frac{p_N}{2}}. 
\end{align}
We take the reflection matrices as $\bar{K}_\pm^B$ in \eqref{eq:K-B2}. 
\begin{align}
    \bar{K}^B_+ = \frac{1}{x-x^{-1}} \begin{pmatrix}
        \L x^{-\frac{\k_+}{2}} & 0 \\ x^{-1}-x & \L x^{\frac{\k_+}{2}}
    \end{pmatrix} ~,~ \bar{K}^B_- = \frac{1}{x-x^{-1}} \begin{pmatrix}
        -\L x^{-\frac{\k_-}{2}} & x^{-1}-x \\ 0 & -\L x^{\frac{\k_-}{2}}
    \end{pmatrix}
\end{align}
The monodromy matrix is
\begin{align}
    T(x)|_{(C^{(1)}_N)^\vee} = \bar{K}^B_+(x,\k_+) t_N(x) \bar{K}^B_-(x,\k_-) t_N^{-}(x).
\end{align}
The spectral curve is
\begin{align}
\begin{split}
    y + \frac{\L^{4N+4}}{y} \frac{1}{(x-x^{-1})^{4}} = & ~ x^N + x^{-N} + \sum_{n=1}^N H_n (x^{N-n}+x^{n-N}) \\
    & - \frac{\L^{2N+2}x}{2(x-1)^2} + (-1)^N \k_+\k_-\frac{\L^{2N+2}x}{2(x+1)^2}
\end{split}
\end{align}
For later convenience, we denote $\k = \k_+\k_-=\pm1$. 
Use a combination of the $SL(2,\BZ)$ action \eqref{eq:sl2z-trans} and birational transformation \eqref{eq:bira-trans} to scale $y \to y(x-x^{-1})^{-2}$ and multiplying the spectral curve by $(x-x^{-1})^2$ gives
\begin{align}
\begin{split}
    & y + \L^{2N+2} \left( x^{\frac{1-(-1)^N\k}{2}} + x^{\frac{(-1)^N\k-1}{2}} \right) +  \frac{\L^{4N+4}}{y} \\
    & = (x-x^{-1})^2 \left[ x^N + x^{-N} + \sum_{n=1}^N (-1)^n H_n (x^{N-n}+x^{n-N}) \right]
\end{split}
\end{align}
This spectral curve coincides with the Seiberg-Witten curve of five dimensional pure $Sp(N)_{\k\pi}$ supersymmetric gauge theory on $\BR^4 \times S^1$ \cite{Brandhuber:1997ua,Zafrir:2015ftn,Hayashi:2023boy,Li:2021rqr}. 

The dimer graph Fig.~\ref{fig:reflective RTC} is the gluing of two $Y^{N,0}$ dimer graphs connected by two type B-2 boundary (Fig.~\ref{fig:B2 graph}). The dimer graph is the same as $Y^{2N+4,0}$ dimer.
The Darboux coordinates of the two $Y^{N,0}$ model are folded \eqref{eq:folding} and related to the canonical coordinates by
\begin{align}
    \xi_{N+1-j}^{-1} = \xi'_{j} = e^{q_j}~,~ \eta_{N+1-j}^{-1} = \eta_{j}' = e^{\frac{p_j}{2}}~,~ j=1,\dots,N.
\end{align}

\paragraph{Mixture}
We are allowed to have a type B-1 boundary \eqref{eq:K-B1} on one end and a type B-2 boundary \eqref{eq:K-B2} on the other. 
The monodromy matrix can be either
\begin{align}
\begin{split}
    T(x) & = \bar{K}^B_+(x,\k_+) t_N(x) K^B_-(x) t^-_N(x), \text{ or } \\
    T(x) & = {K}^B_+ t_N(x) \bar{K}^B_- t^-_N(x).
\end{split}
\end{align}
The spectral curve is given by
\begin{align}
\begin{split}
    & y + \frac{\L^{4N+4}}{y} \frac{x}{(x-x^{-1})^2(x-1)^2} \\
    & = x^N + x^{-N} + \sum_{n=1}^N (-1)^n H_n(x^{N-n} + x^{n-N}) - \frac{\L^{2N+2}}{(x^{\frac12}-x^{-\frac12})^2}.
\end{split}
\end{align}
Scaling $y \to (x-x^{-1})(x-1)y$ and multiplying both sides with $(x-x^{-1})(x-1)$ gives
\begin{align}
\begin{split}
    & y + \L^{2N+2} (x+1) + \frac{\L^{4N+4}x}{y} \\
    & = (x-x^{-1}) (x-1) \left[ x^N + x^N + \sum_{n=1}^N (-1)^n H_n(x^{N-n} + x^{n-N}) \right]
\end{split}
\end{align}
The dimer graph is gluing of two $Y^{N,0}$ dimer through type B1 and B2 boundary (Fig.~\ref{fig:B1 graph} and Fig.~\ref{fig:B2 graph}). The dimer graph shares the same shape as $Y^{2N+3,0}$ dimer model. The Darboux coordinates of the two $Y^{N,0}$ graph obey the folding condition \eqref{eq:folding}.

\subsection{${A}^{(2)}_{2N}$}

\begin{figure}[h!] \centering
\scalebox{1}{
\begin{tikzpicture}[square/.style={regular polygon,regular polygon sides=4},decoration={markings, 
    mark= at position 0.5 with {\arrow[scale=1.3]{stealth}}}
 ]
    
    \node at (-3,-2) [style={circle,draw},thick,inner sep=4pt] (w1) {};
    \node at (-2,-2) [style={circle,draw},thick,inner sep=4pt] (w2) {};
    \node at (-1,-2) [style={circle,draw},thick,inner sep=4pt] (w3) {};
    \node at (1,-2) [style={circle,draw},thick,inner sep=4pt] (w4) {};
    \node at (2,-2) [style={circle,draw},thick,inner sep=4pt] (w5) {};
    \node at (3,-2) [style={circle,draw},thick,inner sep=4pt] (w6) {};

    \draw[thick,double] (w1) -- (w2);
    \draw[thick] (w2) -- (w3);
    \draw[thick] (w3) -- (-.5,-2);
    \draw[thick] (w4) -- (.5,-2);
    \draw[thick] (w4) -- (w5);
    \draw[thick,double] (w5) -- (w6);
    \node at (0,-2) {$\cdots$};

    \node at (2.5,-2) {$\boldsymbol{<}$};
    \node at (-2.5,-2) {$\boldsymbol{<}$};
    
\end{tikzpicture}
}
\caption{Dynkin diagram for $A_{2N}^{(2)}$}\label{fig:A2N-Dynkin}
\end{figure}
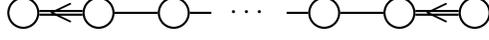

The Dynkin diagram of twisted affine Lie algebra $A_{2N}^{(2)}$ is illustrated in Fig.~\ref{fig:A2N-Dynkin}. The Dynkin diagram has a short root on the left end and a long root on the right. 

The reflection matrix is chosen based on the structure of the Dynkin diagram. The reflection matrix $K^C_-(x)$ is chosen for the long root in \eqref{eq:K-C}. The reflection matrix can be chosen as either $K^B_+(x)$ or $\bar{K}_+^B(x,\k_+)$

The Hamiltonian with with $K^B_+(x)$ for the short root is
\begin{align}
    H_1 = H_A + J^B_+ + J^C_- = H_A + 2\L^2e^{-q_N}\cosh \frac{p_N}{2} + \L^2 e^{2q_1}
\end{align}
The monodromy matrix is
\begin{align}
    T(x) = K_+^B(x) t_N(x) K_-^C t_N^-(x)
\end{align}
The spectral curve is
\begin{align}
\begin{split}
    & y + \frac{\L^{4N+2}}{y} \frac{1}{x-2+x^{-1}}  = x^N + x^{-N} + \sum_{n=1}^{N} (-1)^n H_n (x^{N-n}+x^{n-N}) 
\end{split}
\end{align}
We scale $y \to (x-1)^{-1} y$ and multiply both side of the equation by $(x-1)$: 
\begin{align}
\begin{split}
    & y + \frac{\L^{4N+2}}{xy}   = (x-1)\left[ x^N + x^{-N} + \sum_{n=1}^{N} (-1)^n H_n (x^{N-n}+x^{n-N}) \right]
\end{split}
\end{align}

The Hamiltonian with $\bar{K}^B_+(x)$ for the short root is
\begin{align}
    H_1 = H_A + \bar{J}^B_+ + J^C_- = H_A + \L^2e^{-q_N} e^{\k_+\frac{p_N}{2}} + \L^2 e^{2q_1}
\end{align}
The monodromy matrix is
\begin{align}
    T(x) = \bar{K}_+^B(x) t_N(x) K_-^C t_N^-(x)
\end{align}
The spectral curve is
\begin{align}
\begin{split}
    & y + \frac{\L^{4N+2}}{y} \frac{1}{(x-x^{-1})^2}  = x^N + x^{-N} + \sum_{n=1}^{N} (-1)^n H_n (x^{N-n}+x^{n-N}) 
\end{split}
\end{align}
We scale $y \to (x-x^{-1})^{-1} y$ and multiply both side of the equation by $(x-x^{-1})$: 
\begin{align}
\begin{split}
    & y + \frac{\L^{4N+2}}{xy}   = (x-x^{-1})\left[ x^N + x^{-N} + \sum_{n=1}^{N} (-1)^n H_n (x^{N-n}+x^{n-N}) \right]
\end{split}
\end{align}

The dimer graph is the gluing of two $Y^{N,0}$ models with a type B1 boundary (Fig.~\ref{fig:B1 graph}) or type B2 boundary (Fig.~\ref{fig:B2 graph}) at the end of Fig.~\ref{fig:reflective RTC}. The Darboux coordinates of the two $Y^{N,0}$ dimer graph are subject to the folding in \eqref{eq:folding}. 
It shares the same shape with $Y^{2N+1,0}$ dimer model.
The Darboux coordinates of the two $Y^{N,0}$ model are folded \eqref{eq:folding} and related to the canonical coordinates by
\begin{align}
    \xi_{N+1-j}^{-1} = \xi'_{j} = e^{q_j}~,~ \eta_{N+1-j}^{-1} = \eta_{j}' = e^{\frac{p_j}{2}}~,~ j=1,\dots,N.
\end{align}

\subsection{$B_N^{(1)}$}

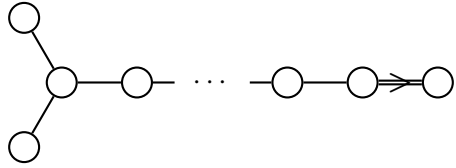
\begin{figure}[h!] \centering
\scalebox{1}{
\begin{tikzpicture}[square/.style={regular polygon,regular polygon sides=4},decoration={markings, 
    mark= at position 0.5 with {\arrow[scale=1.3]{stealth}}}
 ]
    
    \node at (-2.5,-1.14) [style={circle,draw},thick,inner sep=4pt] (w0) {};
    \node at (-2.5,-2.86) [style={circle,draw},thick,inner sep=4pt] (w1) {};
    \node at (-2,-2) [style={circle,draw},thick,inner sep=4pt] (w2) {};
    \node at (-1,-2) [style={circle,draw},thick,inner sep=4pt] (w3) {};
    \node at (1,-2) [style={circle,draw},thick,inner sep=4pt] (w4) {};
    \node at (2,-2) [style={circle,draw},thick,inner sep=4pt] (w5) {};
    \node at (3,-2) [style={circle,draw},thick,inner sep=4pt] (w6) {};

    \draw[thick] (w0) -- (w2);
    \draw[thick] (w1) -- (w2);
    \draw[thick] (w2) -- (w3);
    \draw[thick] (w3) -- (-.5,-2);
    \draw[thick] (w4) -- (.5,-2);
    \draw[thick] (w4) -- (w5);
    \draw[thick,double] (w5) -- (w6);
    \node at (0,-2) {$\cdots$};

    \node at (2.5,-2) {$\boldsymbol{>}$};
    
\end{tikzpicture}
}
\caption{Dynkin diagram for $B_{N}^{(1)}$}\label{fig:B1N-Dynkin}
\end{figure}

The Dynkin diagram of affine Lie algebra $B_{N}^{(1)}$ is illustrated in Fig.~\ref{fig:B1N-Dynkin}. It has a type B boundary on the right end and a type D boundary on the left. 

The reflection matrix is chosen based on the structure of the Dynkin diagram. On the long root the reflection matrix is chosen by $K^D_+(x;q_N,p_N)$ in \eqref{eq:K-D}. The reflection matrix for the short root can be either $K_-^B(x)$ \eqref{eq:K-B1} or $\bar{K}_-^B(x,\k_-)$ in \eqref{eq:K-B2}. The two will be constructed from different dimer graphs. 

The monodromy matrix with reflection matrices $K^D_+$ and $K_-^B$ is 
\begin{align}
\begin{split}
    T(x)|_{B_N^{(1)}} = & ~ {K}^D_+(x;q_{N},p_N) t_{N-1}(x) K_-^B(x) t_{N-1}^-(x) \\
\end{split}
\end{align}
with the spectral curve 
\begin{align}
    y + \frac{\L^{4N+2}}{y} \frac{(x-x^{-1})^2}{x-2+x^{-1}} = x^N + x^{-N} + \sum_{n=1}^{N} (-1)^n H_n (x^{N-n}+x^{n-N}). 
\end{align}
Scaling $Y \to \frac{(x-x^{-1})}{x-1}Y$ and multiply both side of the equation by $(x-1)$ gives 
\begin{align}
    (x-x^{-1}) y + (x-x^{-1})\frac{\L^{4N+2}}{xy} = (x-1) \left[ x^N + x^{-N} + \sum_{n=1}^{N} (-1)^n H_n (x^{N-n}+x^{n-N}) \right]. 
\end{align}

The dimer graph is a gluing of two $Y^{N-1,0}$ dimer graph connected through a type B-1 boundary (Fig.~\ref{fig:B1 graph}) and a type D boundary (Fig.~\ref{fig:double impurity}). 
It shares the same shape to a $Y^{2N-1}$ dimer graph with a type D boundary at the end. The Darboux coordinates of the two $Y^{N-1,0}$ dimer graph are folded \eqref{eq:folding} and related to the canonical variables by 
\begin{align}
    \xi_{N-j}^{-1} = \xi_{j}' = e^{q_{j}}~,~\eta_{N-j}^{-1} = \eta_{j}' = e^{\frac{p_{j}}{2}}~,~ j=1,\dots,N-1~.
\end{align}

\paragraph{}
The monodromy matrix with $K^D_+$ and $\bar{K}_-^B$ reflective boundary is 
\begin{align}
\begin{split}
    T(x)|_{B_N^{(1)}} = & ~ {K}^D_+(x;q_{N},p_N) L^+(x;q_{N-1},p_{N-1}) \cdots L^+_1(x;q_1,p_1)\\
    & \times \bar{K}_-^B(x) L^-(x;q_1;p_1) \cdots L^-(x;q_{N-1},p_{N-1}) \\
\end{split}
\end{align}
with the spectral curve 
\begin{align}
    y + \frac{\L^{4N+2}}{y} \frac{(x-x^{-1})^2}{(x-x^{-1})^2} = x^N + x^{-N} + \sum_{n=1}^{N} (-1)^n H_n (x^{N-n}+x^{n-N}). 
\end{align}
Scaling $y \to \frac{(x-x^{-1})}{(x-x^{-1})}y$ and multiply both side of the equation by $(x-x^{-1})$ gives 
\begin{align}
    (x-x^{-1}) y + (x-x^{-1})\frac{\L^{4N+2}}{y} = (x-x^{-1}) \left[ x^N + x^{-N} + \sum_{n=1}^{N} (-1)^n H_n (x^{N-n}+x^{n-N}) \right]. 
\end{align}
Similar to the $K_-^B$ reflection matrix case, one should not divide $(x-x^{-1})$ on both sides. 
The dimer graph Fig.~\ref{fig:reflective RTC} is a gluing of two $Y^{N-1}$ dimer through a type B2 boundary and a type D boundary. It shares the same shape to a $Y^{2N}$ dimer graph with a type D boundary at the end. The Darboux coordinates of the two $Y^{N-1,0}$ dimer graph are folded \eqref{eq:folding} and related to the canonical variables by 
\begin{align}
    \xi_{N-j}^{-1} = \xi_{j}' = e^{q_{j}}~,~\eta_{N-j}^{-1} = \eta_{j}' = e^{\frac{p_{j}}{2}}~,~ j=1,\dots,N-1~.
\end{align}

\subsection{$(B_N^{(1)})^\vee=A^{(2)}_{2N-1}$}

\begin{figure}[h!] \centering
\scalebox{1}{
\begin{tikzpicture}[square/.style={regular polygon,regular polygon sides=4},decoration={markings, 
    mark= at position 0.5 with {\arrow[scale=1.3]{stealth}}}
 ]
    
    \node at (-2.5,-1.14) [style={circle,draw},thick,inner sep=4pt] (w0) {};
    \node at (-2.5,-2.86) [style={circle,draw},thick,inner sep=4pt] (w1) {};
    \node at (-2,-2) [style={circle,draw},thick,inner sep=4pt] (w2) {};
    \node at (-1,-2) [style={circle,draw},thick,inner sep=4pt] (w3) {};
    \node at (1,-2) [style={circle,draw},thick,inner sep=4pt] (w4) {};
    \node at (2,-2) [style={circle,draw},thick,inner sep=4pt] (w5) {};
    \node at (3,-2) [style={circle,draw},thick,inner sep=4pt] (w6) {};

    \draw[thick] (w0) -- (w2);
    \draw[thick] (w1) -- (w2);
    \draw[thick] (w2) -- (w3);
    \draw[thick] (w3) -- (-.5,-2);
    \draw[thick] (w4) -- (.5,-2);
    \draw[thick] (w4) -- (w5);
    \draw[thick,double] (w5) -- (w6);
    \node at (0,-2) {$\cdots$};

    \node at (2.5,-2) {$\boldsymbol{<}$};
    
\end{tikzpicture}
}
\caption{Dynkin diagram for $(B_{N}^{(1)})^\vee=A^{(2)}_{2N-1}$}\label{fig:A2N2-Dynkin}
\end{figure}
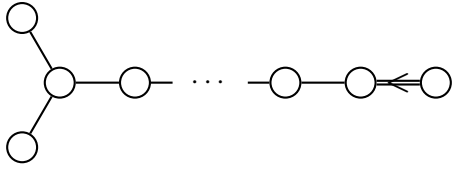

The Dynkin diagram of the dual of affine Lie algebra $B_{N}^{(1)}$, also know as $A^{(2)}_{2N-1}$ twisted affine Lie algebra, is illustrated in Fig.~\ref{fig:A2N2-Dynkin}

The reflection matrices are chosen based on the structure of the Dynkin diagram. 
The monodromy matrix is 
\begin{align}
\begin{split}
    T(x)|_{(B_N^{(1)})^\vee} = & ~ {K}^D_+(x;q_{N},p_N) t_{N-1}(x) {K}_-^C t_{N-1}^-(x).  \\
\end{split}
\end{align}
The twisted matrices are given in \eqref{eq:K-C} and \eqref{eq:K-D}. 
The spectral curve is
\begin{align}
    y +  (x-x^{-1})^2  \frac{\L^{4N}}{y} = x^N + x^{-N} + \sum_{n=1}^N (-1)^N H_n (x^{N-n}+x^{n-N})
\end{align}
Scaling $y \to (x-x^{-1}) y$ gives
\begin{align}\label{eq:BN-dual-spec}
    (x-x^{-1})y +  (x-x^{-1}) \frac{\L^{4N}}{y} = x^N + x^{-N} + \sum_{n=1}^N (-1)^N H_n (x^{N-n}+x^{n-N})
\end{align}
This spectral curve does not coincide with the Seiberg-Witten curve of five-dimensional pure $SO(2N+1)$ supersymmetric gauge theory on $\BR^4 \times S^1$ \cite{Brandhuber:1997ua,Zafrir:2015ftn,Hayashi:2023boy}. 
Instead, the spectral curve of $SO(2N+2)+1\bf{F}$, which can be constructed from the 5-brane web $SU(2N+2) + 10 \bf{F}$. We Higgising between a Coulomb moduli parameters and the fundamental masses
$m_1 = a_{2N+1}=1$, $m_2=a_{2N+2}=i\pi$. 
\begin{align}
    y + \frac{(x-x^{-1})^4 \times (x-x^{-1})^2\L^{4N}}{y} = (x-x^{-1})^2 \left[ x^N + x^{-N} + \sum_{n=1}^N H_n (x^{N-n}-x^{n-N}) \right]
\end{align}
Scaling $y\to (x-x^{-1})^2 y$ and dividing both side with $x-x^{-1}$ recovers \eqref{eq:BN-dual-spec}. 
On the 5-brane web picture of the supersymmetric gauge theory, a color D5-brane and a flavor D5-brane are brought to the $O7^+$ plane and combined with their reflection. 
After the Higgsing takes place, a D5 brane can be pulled away from the 5-brane web, leaving a pure $SO(2N+1)$ where a half D5-brane is stuck at the point of an $O7^+$-plane. See Fig.~7 in \cite{Hayashi:2023boy}

The dimer graph Fig.~\ref{fig:reflective RTC} is a gluing of two $Y^{N-1}$ dimer with a type D boundary at the end. It shares the same shape to a $Y^{2N-2,0}$ dimer graph with a type D boundary at the end. The Darboux coordinates of the two $Y^{N-1,0}$ dimer graph are folded \eqref{eq:folding} and related to the canonical variables by 
\begin{align}
    \xi_{N-j}^{-1} = \xi_{j}' = e^{q_{j}}~,~\eta_{N-j}^{-1} = \eta_{j}' = e^{\frac{p_{j}}{2}}~,~ j=1,\dots,N-1~.
\end{align}

\subsection{$D^{(1)}_N$}

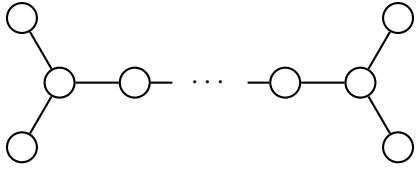
\begin{figure}[h!] \centering
\scalebox{1}{
\begin{tikzpicture}[square/.style={regular polygon,regular polygon sides=4},decoration={markings, 
    mark= at position 0.5 with {\arrow[scale=1.3]{stealth}}}
 ]
    
    \node at (-2.5,-1.14) [style={circle,draw},thick,inner sep=4pt] (w0) {};
    \node at (-2.5,-2.86) [style={circle,draw},thick,inner sep=4pt] (w1) {};
    \node at (-2,-2) [style={circle,draw},thick,inner sep=4pt] (w2) {};
    \node at (-1,-2) [style={circle,draw},thick,inner sep=4pt] (w3) {};
    \node at (1,-2) [style={circle,draw},thick,inner sep=4pt] (w4) {};
    \node at (2,-2) [style={circle,draw},thick,inner sep=4pt] (w5) {};
    \node at (2.5,-1.14) [style={circle,draw},thick,inner sep=4pt] (w6) {};
    \node at (2.5,-2.86) [style={circle,draw},thick,inner sep=4pt] (w7) {};

    \draw[thick] (w0) -- (w2);
    \draw[thick] (w1) -- (w2);
    \draw[thick] (w2) -- (w3);
    \draw[thick] (w3) -- (-.5,-2);
    \draw[thick] (w4) -- (.5,-2);
    \draw[thick] (w4) -- (w5);
    \draw[thick] (w5) -- (w6);
    \draw[thick] (w5) -- (w7);
    \node at (0,-2) {$\cdots$};

    
\end{tikzpicture}
}
\caption{Dynkin diagram for $D_{N}^{(1)}$}\label{fig:D-Dynkin}
\end{figure}

To complete the story, we shall mention the $D^{(1)}_N$ RTC, which is discussed in detail in \cite{Lee:2024bqg}. The Dynkin diagram of $D^{(1)}_N$ is illustrated in Fig.~\ref{fig:D-Dynkin}

The Hamiltonian is 
\begin{align}
    H_1 = H_A + 2 \L^2 e^{q_1+q_2} \cosh \frac{p_1-p_2}{2} + 2\L^2 e^{-q_{N-1}-q_N} \cosh \frac{p_{N-1}-p_N}{2} + \L^4 e^{2q_2} + \L^4 e^{-2q_{N-1}}~.
\end{align}
The monodromy matrix is 
\begin{align}
\begin{split}
    T(x)|_{(D_N^{(1)})^\vee} = & ~ {K}^D_+(x;q_{N},p_N) t_{N-2}(x) {K}_-^D(x;q_1,p_1) t_{N-2}^-(x) \\
\end{split}
\end{align}
with reflection matrices $K_\pm^D$ given in \eqref{eq:K-D}. 
The spectral curve is
\begin{align}
    y +  (x-x^{-1})^4  \frac{\L^{4N}}{y} = x^N + x^{-N} + \sum_{n=1}^N (-1)^N H_n (x^{N-n}+x^{n-N})~.
\end{align}
Scaling $y \to (x-x^{-1})^2  y$ gives
\begin{align}
    (x-x^{-1})^2 y +  (x-x^{-1})^2 \frac{\L^{4N}}{y} = x^N + x^{-N} + \sum_{n=1}^N (-1)^N H_n (x^{N-n}+x^{n-N})
\end{align}
This spectral curve coincides with the Seiberg-Witten curve of five dimensional pure $SO(2N)$ supersymmetric gauge theory on $\BR^4 \times S^1$ \cite{Hayashi:2023boy,Nekrasov:1996cz}. 
The dimer graph Fig.~\ref{fig:reflective RTC} is a gluing of two $Y^{N-2,0}$ dimer graph through two type D boundary. The Darboux coordinates of the $Y^{N-2,0}$ models are related to the canonical coordinates by
\begin{align}
    \xi_{N-1-j}^{-1} = \xi_{j}' = e^{q_{j+1}}~,~\eta_{N-1-j}^{-1} = \eta_{j}' = e^{\frac{p_{j+1}}{2}}~,~ j=1,\dots,N-2~.
\end{align}

\section{Summary}\label{sec:summary}

In this note, we construct the dimer graph for the relativistic Toda chains (RTC) defined on several Lie algebras $\fg$. The construction is by gluing two open type A RTC with a proper graph for the reflective boundaries. The choice of the reflective boundary depends on the structure of the Lie algebra's Dynkin diagram. 
Our work presents extra evidence that RTC defined based on semi-simple Lie group are cluster integrable systems. 
The Lax matrices and reflection matrices of the RTC are constructed by considering the Baker-Akhiezer function of the dimer graph's Kasteleyn matrix. 

We end this note with some future directions. 
\begin{itemize}
    \item An RTC can be defined based on any Lie algebra, including the exceptional Lie algebras E, F, and G. In particular, the exceptional $G_2$ can be obtained by folding from $D_4$ or $B_3$. An immediate question is whether the folding on the Lie algebra level extends to the RTC.  

    \item There are three RTCs we mention defined on $\fg = C_N^{(1)}$, $B_{N}^{(1)}$, and $A_{2N}^{(2)}$ whose dual groups are the twisted Lie groups. \\
    5d $\CalN=2$ super Yang-Mills theory with twisted Lie groups are constructed from the 6d $\CalN=(2,0)$ theory with an outer-automorphism twist on the compactified circle \cite{Tachikawa:2011ch,Duan:2021ges}. \\
    In principle, 5d $\CalN=1$ super Yang-Mills theories with twisted Lie groups can be obtained by deforming the 5d $\CalN=2$ super Yang-Mills then integrating out the adjoint mass. We want to know whether the Seiberg-Witten curves of super Yang-Mills with twist Lie algebras match with the spectral curve of the RTCs.   
    
    \item The X-cluster algebra $\CalX_\S$ has a natural quantization $\CalO_q(\CalX_\Sigma)$. 
    The Lax operators of the quantum cluster integrable systems obey the Yang-Baxter equation and the reflection matrices obey a quantum version of \eqref{eq:reflect-classical}. Quantization conditions for type A RTC has been studied via Bethe/Gauge correspondence in analogous to 4d  \cite{Nikita:V,Nikita-Shatashvili-1,Nikita-Shatashvili-2,Nikita-Shatashvili-3,Nikita-Pestun-Shatashvili}. However, the Nekrasov-Shatashvili free energy, which worked perfectly in 4d, is insufficient in the 5d. The correct quantization requires towers of non-perturbative effect addressed by introducing Wilson loop via topological string \cite{Hatsuda:2015qzx,Franco:2015rnr,Grassi:2014zfa,Grassi:2017qee}. \\ 
    The quantum Hamiltonians and wave function are obtained from co-dimensional two monodromy defect \cite{Bonelli:2022iob,Bullimore:2014awa,Chen:2019vvt,Chen:2020rxu,Lee:2020hfu,Lee:2024jae,Jeong:2024mxr}, effectively coupling a 3d $\CalN=2$ quiver gauge theory to the 5d $\CalN=1$ gauge theory. \\
    The Baxter $Q$-operator is constructed in the gauge theory by co-dimensional two canonical defect \cite{Jeong:2021rll,Jeong:2023qdr,Jeong:2025yys}. An interesting observation is that for type A open RTC the Baxter $Q$-operator is realized by a series of cluster mutation \cite{schrader2018b}. It is natural to ask if the cluster mutation construction of $Q$-operator can be extended to other cluster integrable systems.
    
    \item For some integrable chains special kind of duality can be observed on both the classical and quantum level. A system with $N$-dimensional auxiliary space on $M$ sites shares the Hamiltonians with some other system with $M$-dimensional auxiliary space on $N$ sites. Under the duality the spectral parameters that the monodromy operator depends on, and the spectral parameter of the characteristic equation exchange. Hence this duality are often called \emph{spectral duality}, sometimes also referred as \emph{level-rank duality}. \\
    For cluster integrable chain on a dimer, the spectral duality can manifest itself as rotation of the dimer graph by 90 degrees, sometimes with a twist. It is known a type A RTC with $N$ sites is spectral dual to a $\mathfrak{gl}_1$ chain with $N$ sites with a cyclic twist \cite{Marshakov:2019vnz}. \\
    On the supersymmetric gauge theory side, this transformation turns the theory of a $SU(N)$ hypermultiplets with only $SU(N) \times SU(N)$ flavor symmetry to pure $SU(N)$ gauge theory. \\
    We would like to know if other RTCs have spectral duality analogous to type A. In particular, whether the duality can be observed on the dimer graph level if it exists.   
    
\end{itemize}

\paragraph{Conflict of interest statement}

There are no conflicts of interest. 

\paragraph{Data availability statement}

There are no additional data. 

\appendix


\bibliographystyle{utphys}
\bibliography{SU-Sp}

\end{document}